\documentclass[pra,twocolumn,aps,showpacs]{revtex4-1}

\usepackage{amsmath}
\usepackage{amssymb}
\usepackage{braket}
\usepackage{times}
\usepackage{hyperref}
\usepackage{graphicx}
\usepackage{color}
\usepackage{bm}
\everymath{\displaystyle}
\usepackage{tikz}
\usetikzlibrary{decorations.markings}
\usetikzlibrary{decorations.pathmorphing,snakes}

\begin{document}

\title{Condensates in double-well potential with synthetic gauge potentials and
       vortex seeding }

\author{Rukmani Bai}
\affiliation{Physical Research Laboratory,
             Navrangpura, Ahmedabad-380009, Gujarat,
             India\\ Indian Institute of Technology,
             Gandhinagar, Ahmedabad-382424, Gujarat, India}
\author{Arko Roy}
\affiliation{Physical Research Laboratory,
             Navrangpura, Ahmedabad-380009, Gujarat,
             India\\ Max-Planck-Institut f{\"u}r Physik komplexer Systeme,
           N{\"o}thnitzer Stra\ss e 38, 01187 Dresden, Germany}
\author{D. Angom}
\affiliation{Physical Research Laboratory,
             Navrangpura, Ahmedabad-380009, Gujarat,
             India}
\author{P. Muruganandam}
\affiliation{ School of Physics,
              Bharathidasan University, Tiruchirapalli 620 024,
              Tamil Nadu, India}

\date{\today}


\begin{abstract}

We demonstrate an enhancement in the vortex generation when artificial gauge 
potential is introduced to condensates confined in a double well potential. This
is due to the lower energy required to create a vortex in the low condensate 
density region within the barrier. Furthermore, we study the transport of 
vortices between the two wells, and show that the traverse 
time for vortices is longer for the lower height of the well. We also 
show that the critical value of synthetic magnetic field to inject 
vortices into the bulk of the condensate is lower in the double-well 
potential compared to the harmonic confining potential. 
\end{abstract}


\maketitle

\section{Introduction}

   Charged particles experience  Lorentz force in the presence of 
magnetic fields, and in condensed matter systems, it is the essence
for a host of fascinating phenomena like the integer quantum Hall 
effect \cite{klaus_86,yennie_87}, fractional quantum Hall effect 
\cite{stormer_99a,stormer_99b}, and the quantum spin Hall 
effect~\cite{konig_08}. In contrast, the dilute quantum gases of atoms, which 
have emerged as excellent proxies of condensed matter systems, do not 
experience Lorentz force as these are charge neutral. This can, however, be 
remedied with the creation of artificial gauge fields through laser 
fields~\cite{juzeliunas_04,zhu_06,spielman_09,karina_12,goldman_14}. Thus, 
with the artificial gauge potentials it is possible to explore phenomena such 
as the quantum Hall effect, and the quantum spin Hall effect~\cite{zhu_06} in 
dilute atomic quantum gases. The introduction of synthetic magnetic field 
arising from artificial gauge field is also an efficient approach to 
generate quantized vortices in Bose-Einstein condensates (BEC) of dilute
atomic gases. This method has the advantage of having time-independent 
trapping potentials over the other methods like rotation 
~\cite{madison_00,shaeer_01,haljan_01}, topological phase 
imprinting~\cite{isoshima_00,lea_02}, or phase engineering in 
two-species condensates\cite{williams_99,matthews_99}. In addition, it has
the possibility to inject large ensembles of vortices, and an efficient
scheme to nucleate vortices with synthetic magnetic fields was demonstrated
in a recent work~\cite{price_16}. The method relies on the 
creation of an inhomogeneous synthetic magnetic field, which has its
maxima coincident with the low density region of a spatially separated 
pair of BECs.

  In this work we examine a scheme to nucleate vortices in BECs 
through synthetic magnetic field by employing the density gradient associated 
with a double well trapping potential. The advantages of the scheme
are: vortices are generated in the bulk of the condensate; shorter relaxation
time after nucleation; and higher density of  vortices. In contrast, the other 
methods like rotating traps and phase imprinting nucleates vortices at the
periphery. These then migrate to the bulk and as the process is diabatic, the 
relaxation times are long. Hence, the present scheme is better suited to 
explore phenomena associated with high vortex densities 
like quantum turbulence \cite{tsubota_13,nemirovskii_13}. BECs in double 
well potentials were 
first theoretically studied to examine the physics of Josephson currents 
\cite{juha_86,dalfovo_96,smerzi_97}, latter observed in experiments
\cite{cataliotti_01,albiez_05,levy_07,trenkwalder_16}, and 
studied numerically in a recent work\cite{garcia}. For our study, we 
theoretically consider the case of a double well potential which is engineered 
from a harmonic potential by introducing a Gaussian barrier. For alkali metal 
atoms the barrier is a blue-detuned light sheet obtained from a laser beam, 
and  such setups have been used in experiments to observe the matter wave 
interference \cite{andrews_97}, Josephson effects \cite{levy_07} and collision 
of matter-wave solitons \cite{nguyen_14}. The artificial gauge potential is
introduced through Raman coupling \cite{spielman_09}, and as a case study 
we consider the case of $^{87}$Rb BEC.  We use Gross-Pitaevskii (GP) equation 
for a mean-field description of the BEC with the artificial gauge potential. 
In this work we quench the artificial gauge potential by increasing the Raman 
detuning, and simultaneously increase the height of the barrier potential. 
It is found that the extended low density region associated with the barrier 
promotes the formation of vortices. However, the quench imparts energy to the 
BEC and transfer it to an excited state. For comparison, we 
also examine the vortex generation in the case of uniform BEC \cite{Gaunt_13}. 
Such a system, devoid of trap induced inhomogeneities, is better for 
quantitative comparison of experimental results with theory. This was 
demonstrated in a recent study on wave turbulence in uniform BECs 
\cite{navon_16}. To induce relaxation of the condensate to the ground state, 
we use the standard approach of introducing a dissipative 
term \cite{choi_98,tsubota_02,yan_14}. The presence of the 
dissipative term in the GP equation is consistent with the experimental 
observations of dissipation or damping \cite{mewes_96, jin_97}, which arises 
from the interaction between the condensate and non-condensate atoms. 

The paper is organized as follows, in Section. II we provide a description of
the theory on how to generate artificial gauge potential in BECs using Raman
coupling. Then, we incorporate the gauge potential in the Gross-Pitaevskii
equation to arrive at a mean field description of BEC. In Section. III, we 
present the results of numerical computations, and discuss the implications.
We, then, conclude with the key observations. 

%

\section{BEC in artificial gauge potentials}

To study the vortex formation in double well with synthetic magnetic field in 
BECs theoretically, we consider the scheme based on light induced gauge 
potential proposed in ref. \cite{spielman_09}. In particular, we 
consider a quasi-2D BEC along the $xy$-plane of two level atoms, which in the 
present work is taken as the $F=1$ ground state of $^{87}$Rb atoms. To generate 
spatial inhomogeneity an external magnetic field $B(y) = B_0 + \Delta B(y)$ is
applied along the $y$ direction. Here $B_0$ is the static magnetic field which 
introduces a linear Zeeman splitting of the ground state manifold. The energy 
levels are separated by 
$\Delta_{\rm z} = g \mu_{\rm B} B_{\rm0}$, and 
$\delta(y) = g \mu_{\rm B} \Delta B(y)$ is the measure of detuning from Raman 
resonance. The constants  $g$ and $\mu_{\rm B}$ are the atomic Land\'{e} factor
and Bohr magneton, respectively. The two levels in the ground state are
Raman coupled through two counter-propagating laser beams 
passing through the BEC along $\pm x$ directions~\cite{lin_09}. The momentum 
transferred to the atoms through interactions with the Raman lasers induces a 
change in the kinetic energy part of the Hamiltonian through the vector 
potential term $A_x$. The modified Hamiltonian, however, remains gauge 
invariant, and there is a corresponding synthetic magnetic field 
$B_z=-\partial A_x/\partial x$. 

%
%
\subsection{Modified Gross-Pitaevskii equation}
In the absence  of Raman coupling, the Hamiltonian of the quasi-2D BEC 
confined in an harmonic trapping potential ${\hat V}_{\rm trap}$ is
\begin{equation}
  \hat{H} = {\hat H}_x + {\hat H}_y + {\hat V}_{\rm trap} + {\hat H}_{\rm int},
  \label{gp.eq}
\end{equation}
where ${\hat H}_x, {\hat H}_y$ represent the kinetic energy part of the 
Hamiltonian term along $x, y$ directions respectively, and 
${\hat H}_{\rm int}$ denotes the interaction energy between the atoms.
Let $\ket{1} = \ket{1,0}$ and $\ket{2} = \ket{1,-1}$ denote the two states in 
the ground state manifold of the atoms. The Raman lasers are along the $x$ 
direction, and hence, the addition of the atom-light coupling term modifies
$H_x$ to 
\begin{eqnarray}
  {\hat H}_{x}= E_{\rm r}
  \begin{pmatrix}
  \left(\tilde{k}_{x}+1\right)^2 -\frac{\hbar \delta}{2E_{\rm r}} && 
  \frac{\hbar \Omega}{2 E_{\rm r}}\\\\
  \frac{\hbar \Omega}{2  E_{\rm r}} &&  \left(\tilde{k}_{x}-1\right)^2
  +\frac{\hbar \delta}{2 E_{\rm r}} 
  \end{pmatrix},
  \label{modhx}
\end{eqnarray}
where $E_{\rm r} = \left(\hbar^2 k_{\rm r}^2/2 m\right)$ is the recoil
energy, and $\tilde{k}_{x} =(k_{x}/k_{\rm r})$ with $k_{x}$ as the
$x$-component of the wave-vector, $\Omega$ is the Raman coupling between
two levels, and $\delta$ is the Raman detuning. 

 To derive the modified Gross-Pitaevskii (GP) equation, we diagonalize 
${\hat H}_{x}$ and obtain the dispersion relation for the two 
levels in the limit of strong Rabi coupling, $\hbar\Omega\gg 4 E_r$. This 
ensures that there is single energy minima of the system and 
lead to the following: there is a change in the 
momentum along the $x$ direction which provide a gauge potential 
$e A_{x}/\hbar k_{\rm r} = \tilde{\delta}/ (\tilde{\Omega} \pm4)$ 
in the system; and from the light-atom coupling the atoms acquires an
effective mass $m^*$ defined by $m^{*}/m =\tilde{\Omega}/(\tilde{\Omega} 
\pm 4)$. Here $\pm$ denotes the two energy levels in the system and
$\tilde{\delta} = \hbar \delta/E_r$, $\tilde{\Omega} = \hbar \Omega/E_r$.
Based on this Hamiltonian and restricting the dynamics to only 
the lowest dressed state, the behaviour of such a 
condensate in the presence of artificial gauge fields is governed by the
following dimensionless modified Gross-Pitaevskii (GP) equation 
\begin{eqnarray}
  \label{scale_gp}
  i\frac{\partial \phi(x,y,t)}{\partial t} &=& 
  \Bigg[-\frac{1}{2}\frac{m}{m^*}\frac{\partial^2}{\partial x^2} - \frac{1}{2}
  \frac{\partial^2}{\partial y^2} 
  + \frac{i 2\pi \delta^{'}}{\lambda_{\rm L} \Omega E_{\rm r}} y\frac{\partial}
  {\partial x}  \nonumber\\&&\hspace{-1.5cm} 
  + \frac{1}{2} x^2 + \frac{1}{2} y^2 \left(1 
  + \frac{2 C_{\rm rab} \delta^{'2}}{E_{\rm r}}\right)
  + g_{\rm 2D}\vert \phi(x,y,t)\vert^2 \nonumber\\&&\hspace{-1.5cm}
  - \left(\frac{\Omega - 2}{2}\right) E_{\rm r} \Bigg]\phi(x,y,t).
\end{eqnarray}
In the above equation, all the parameters having the dimensions of length, 
energy, and time have been scaled with respect to the oscillator 
length $a_{\rm osc} = \sqrt{\hbar/m \omega_x}$, energy $\hbar
\omega_x$ and time $\omega_x t$ respectively. For simplicity of notations,
from here on we will represent the transformed quantities 
($\Omega, \delta^{'}, E_{\rm r}, \lambda_{\rm L}$) without tilde. The 
condensate wavefunction is represented by $\phi(x,y,t)$, 
$C_{\rm rab} = \left(1/\Omega(\Omega - 4) + (4 - \Omega)/4(\Omega + 
4)^2\right)$, $\delta = \delta^{'} y$, $\delta^{'}$ is defined to be the 
detuning gradient, 
$\Omega$ is the Rabi frequency, $E_{\rm r} = \left(2\pi^2/\lambda_{\rm L}^2
\right)$ is the recoil energy of electrons, 
$g_{\rm 2D} = 2 a_s N \sqrt{2\pi \lambda}/a_{\rm osc}$ 
is the interaction energy with $N$ as the total number of atoms in the 
condensate, and $\lambda \gg 1$ is the trap anisotropy parameter along the 
$z$ direction.

\subsection{Double Well (DW) Potential}
For the present work, we consider quasi-2D BEC confined in a double well 
potential 
\begin{equation}
  \label{trp_bar}
  V_{\rm dw} = V_{\rm trap} + U_{0}{\rm exp}{(-2y^2/\sigma^2)},
\end{equation}
where $U_{0}$ and $\sigma$ are the depth and width of the double well 
potential respectively and $ V_{\rm trap}= (1/2)m\omega_\perp^2(x^2 + y^2)$ is 
the harmonic potential along $x$ and $y$ directions, and we have considered
the symmetric case $\omega_\perp=\omega_x=\omega_y$.

The presence of the double well potential modifies the density distribution, 
breaks the rotational symmetry of the condensate, and brings about novel 
effects in the dynamical evolution of the condensate which forms the main 
topic of the present study.


\subsection{Thomas Fermi Correction in the condensates density}

 The focus of the present work, as mentioned earlier, is to examine the 
formation of vortices in the quasi-2D BEC with the introduction of artificial 
gauge potential. It has been shown in previous works on rotated condensates 
that vortices are seeded at the periphery of the condensate cloud, where the
low density of the condensate is energetically favourable for the formation of 
vortices \cite{hodby_01}. This is due to the presence of nodeless surface 
excitations \cite{khawaja_99}, which create instabilities in the condensate and 
lead to the nucleation of vortices \cite{simula_02}. At later times the 
vortices migrate and enter the bulk of the condensate. To analyse the density 
distribution and optimal conditions for generation of vortices, consider a BEC 
with large number of particles. The condensate is then well described with the 
Thomas-Fermi (TF) approximation as the interaction energy dominates in the GP 
equation, and the kinetic energy can be neglected in the bulk of the 
condensates as the spatial gradient is negligible. In this approximation 
the density in the bulk is
$ \vert\phi_{\rm TF}\vert^2 = (\mu - V_{\rm trap})/(2 g_{2D}),$
where $\mu$ is the chemical potential of the condensate.
The TF approximation, however, fails at the periphery of condensate as the 
$\phi_{\rm TF}$ is discontinuous at the boundary \cite{pethick_08}. However, 
vortices are seeded at the peripheral regions where the TF approximation may 
break down. Similar conditions are applicable to the densities at the 
edges of the double well potential considered in the present work. For this  we 
take $\mu$ in terms of TF radius $R$ and trapping potential in term of radial 
distance $r$. The correction to the TF density at the edges, similar to the
harmonic trapping potential, is 
\begin{equation}
  \label{TF_c}
  \vert\phi_{\rm TF}^c\vert^2 = \frac{R^2 - r^2}{2 g_{2D}}\Big[1 - 
  \frac{R^2}{2(R^2 - r^2)^3}\Big]^2.
\end{equation}
In the above equation, first term is TF density in the bulk of condensate 
and second term is correction to the TF density. Now, the density at the 
boundary is calculated as $n_c = \vert\phi_{\rm TF}\vert^2 -
\vert\phi_{\rm TF}^c\vert^2$, and for the region within the barrier of the 
double well, TF approximation is valid as the barrier potential decays 
exponentially. Accordingly, the density distribution is
\begin{equation}
  \vert\phi_{{\rm TF}}^b\vert^2 = \frac{R^2 - r^2 - U_{0}{\rm exp}
  {(-2y^2/\sigma^2)}}{2 g_{2D}},
\end{equation}
here $U_0$ and $\sigma$ are depth and width of the double well potential. The 
above density distribution is symmetric about the $x$-axis, and hence, the 
low density region is more extended compared to the peripheral region of a 
harmonic trapping potential. So, the density variation arising from  the 
potential barrier enhances the formation of vortices.


\section{Results and discussion}

\subsection{Numerical details}
For the present study, we numerically solve the Gross-Pitaevskii equation in 
imaginary time at zero temperature in the absence of the artificial gauge
potential, which is equivalent to setting $\delta^{'}$ and $\Omega$
to zero. For this we use the split-step Crank-Nicolson method 
\cite{muruganandam_09,dusan_12,sataric_16,young_16} and the solution 
obtained is the equilibrium ground state. To dynamically evolve the 
condensate, we propagate the stationary state solution in real time using 
Eq.~(\ref{scale_gp}). Furthermore, we introduce the artificial gauge potential 
by varying $\delta^{'}$  from $0$ to 
$ 3 \times 10^9$ Hz/m within $\approx 202$ ms, but the value of $\Omega$ is 
kept constant throughout the evolution. Afterwards the system is evolved 
freely for up to $t\approx 962$ ms when it relaxes to a steady state. For the
present work we consider $^{\rm 87}\rm Rb$ condensate with $N = 10^5$ atoms, 
and the $s$-wave scattering length is $a_s = 99 a_0$. The trapping potential
parameters are chosen to be $\omega_x = \omega_y = 2 \pi \times 20$ Hz, and 
$\lambda = 40$ which satisfies the quasi-2D condition. The Raman lasers
considered for our calculations have wavelength $\lambda_{\rm L} = 790$ nm.
The Rabi frequency is taken to be $\Omega = 6E_r$, where $E_r$
is scaled with $\hbar \omega_x$. This choice of parameters is consistent with 
the experimental setting of Spielman \emph{et al.}\cite{spielman_09}.

\begin{figure}
  \begin{center}
  \includegraphics[width=8.5cm]{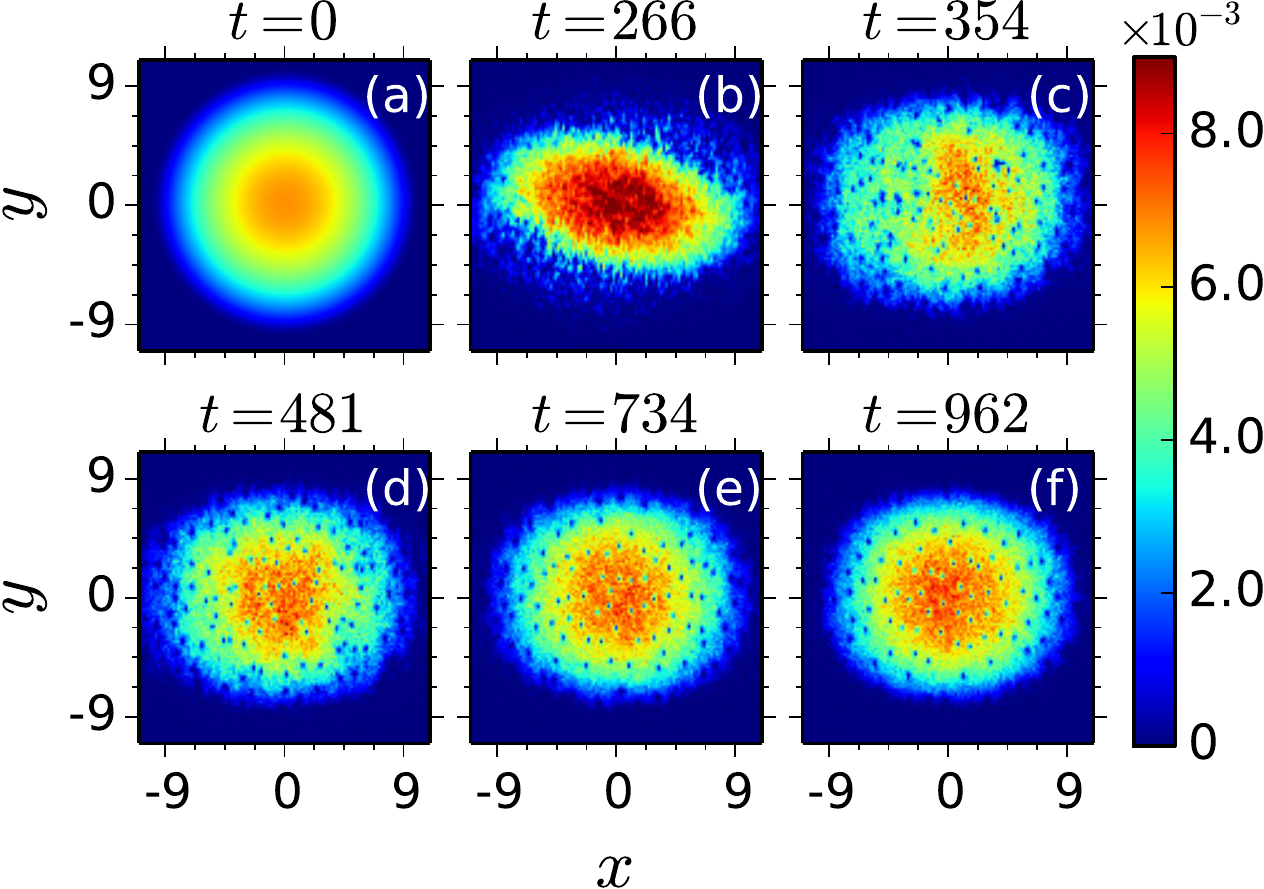}
  \caption{Generation of vortices in the absence of dissipation. The time (in
  units of ms) is shown above the plots. Here $x$ and $y$ are measured in units 
  of $a_{\rm osc}$. Density is measured in units of $a_{\rm osc}^{-2}$ and is 
  normalized to unity.}
  \label{u0_0}
  \end{center}
\end{figure}

\begin{figure}
  \begin{center}
  \includegraphics[width=8.5cm]{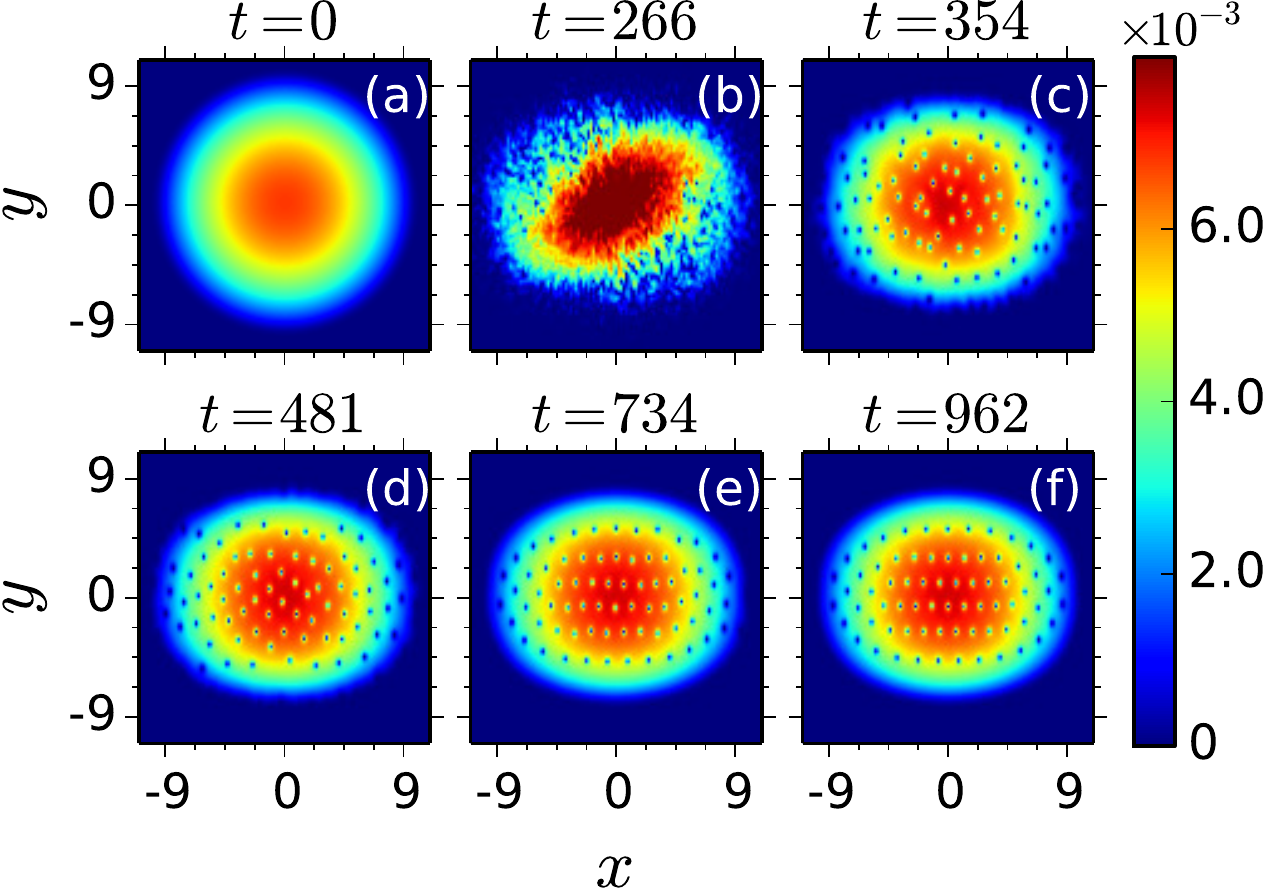}
  \caption{Generation of vortices in the presence of dissipation. The time
  (in units of ms) is shown above the plots. Here $x$ and $y$ are measured in 
  units of $a_{\rm osc}$. Density is measured in units of $a_{\rm osc}^{-2}$ 
  and is normalized to unity.}
  \label{u0_0_g}
  \end{center}
\end{figure}


\subsection{Harmonic potential}
At the start of the real time evolution, or beginning of the dynamical 
evolution $t=0$, the condensate is rotationally symmetric, and is devoid of any 
topological defects. This is shown in Fig.~\ref{u0_0}(a). As the artificial 
gauge potential is switched on in by introducing $\delta^{'}$ with constant
$\Omega$, the rotational symmetry of the condensate is broken
since the effective frequencies along $x$ and $y$ directions are unequal 
due to the term $2C_{\rm rab}\delta'^{2}/E_{\rm r}$ in Eq.~(\ref{scale_gp}). 
The condensate thus departs from being circularly symmetric and acquires an 
elliptic structure, which is discernible from the density profiles shown in 
Fig.~\ref{u0_0}. Furthermore, the angular-momentum like term 
$i(2\pi\delta')/(\lambda_{\rm L}\Omega E_{\rm r})(y\partial\phi(x,y,t) )/
(\partial x) $ in Eq.~(\ref{scale_gp}) is non-zero and induces a deformation to 
the condensate. The combined effects of these two effects and  an increase in the 
energy of the system favour the seeding of topological defects or vortices in the 
condensate. Initially, the vortices are generated at the periphery, where the 
density is low and fluctuations in phase are more prominent, and at later times
the vortices migrate to the bulk of the condensate. As the system relaxes 
towards a steady state, the vortices acquires a spatially disordered distribution 
to minimize the total energy. The nature of the spatial distribution of vortices 
implies that the system is in a higher energy state, and  this is evident 
from the vortex distribution as shown in Fig.~\ref{u0_0}(f).
To include the effects of dissipation which may be present due to quantum
and thermal fluctuations, or due to loss of atoms from the trap because of 
inelastic collisions in the condensate we add the dissipative term 
$-\gamma\partial \phi(x,y,t)/\partial t$ in Eq.~(\ref{scale_gp}) 
and examine the dynamical evolution of the condensate. Here, we set
$\gamma = 0.003$ based on the results from previous work~\cite{yan_14}. This 
leads to loss of energy from the condensate and the
condensate dynamically evolves to it's ground state. As a consequence the
vortices self organise into a vortex lattice and  the evolution towards the
vortex lattice is as shown in Fig.~\ref{u0_0_g}.

%
\subsection{Double well potential}

To study the dynamics of vortex generation and their transport in double well
potential we solve time dependent GP Eq.~(\ref{scale_gp}) with the potential
given in Eq.~(\ref{trp_bar}). Like in the previous case, we include a 
dissipative term to allow the system to relax to it's ground state, which is 
with a vortex lattice. For the numerical computation, we take the width of the 
barrier in the double well potential as $\sigma = 0.7 \mu$m. To obtain the 
initial state, like in the previous case, we again consider imaginary time 
ground state solution of a quasi-2D BEC of $^{\rm 87}\rm Rb$ atoms without 
$\delta '$, $\Omega$, and $U_0$. We, then evolve the solution in real time as 
described earlier. During the evolution in real time, we ramp up or quench the 
value of $\delta'$ and $U_0$, but keeping $\Omega$ fix. Increment in $\delta'$ 
introduces artificial gauge potential in the system, and vortices are generated 
with time. We find that the double well potential has enhanced vortex formation.
The vortices are generated in the barrier region between the two wells as it is
a region of low density, and the energy per vortex is lower in this regime.

 The enhancement of vortex formation in the double well potential can be 
understood in terms of the excitation energy of a single vortex. In the case
of harmonic potential the energy of the vortex located at a radial distance
$b$ from the center in the TF approximation is $\epsilon_v \simeq (4/3)\pi R_z 
n(0)(\hbar^2/m){\rm ln}(R/\xi_0)(1- b^2/R^2)^{3/2}$. Where, $n(0)$ is the 
density at the center, when vortex is not present, $R$ is the Thomas-Fermi 
radius, $R_z$ is the length along $z$ direction and $\xi_0$ is the healing 
length \cite{pethick_08}. For the quasi-2D system $R_z$ can be evaluated using 
the anisotropy parameter $\lambda$ and $\mu$ in the TF approximation. 
Based on this expression we find that the energy of a vortex at the center of
the condensate with only the harmonic trapping potential or without the
barrier is 0.028 $\hbar \omega_x$, which is lower than the value of 
0.094 $\hbar \omega_x$ obtained from the numerical results. The difference 
could be due to deviation from the TF approximation. From the numerical results, 
without the barrier, the energy of a vortex located at a radial distance of 
9.0 $a_{\rm osc}$ is 0.008 $\hbar \omega_x$. Here, the radial distance 
considered correspond to the peripheral region where vortices first appear. 
In the case of double well potential the energy of a vortex at the center of 
the barrier and at the same radial distance is 0.007 $\hbar \omega_x$, which is 
lower than the previous case. In terms of absolute values the energy difference 
is not large, but as discussed latter, the presence of the barrier in the double
well makes a significant difference in the dynamical evolution and generation of
vortices. Since we quench two parameters of the system, $\delta'$ and $U_0$, we 
examine the system in terms of the relative quench rates. For this we define 
$R_1= \lambda_{\rm L} \partial\delta '/\partial t$  and 
$R_2 = \partial U_0/\partial t$ as the quench rate of the artificial gauge 
potential, and the barrier height between the two wells, respectively.
Where $\lambda_{\rm L}$, $\delta'$, and $U_0$ are the dimensionless quantities, 
and $\delta '$ and $U_0$ are ramped within a period of $t = 202$ ms. The value 
of $\delta'$ vary from $0$ to $3 \times10^9$ Hz/m as defined earlier. Here, 
$R_1$ affects the vortex generation, and $R_2$ affects the transport of vortices
between the two wells. We consider three cases, depending on the relative 
values of $R_1$ and $R_2$.

%

\begin{figure}
  \begin{center}
  \includegraphics[width=8.5cm]{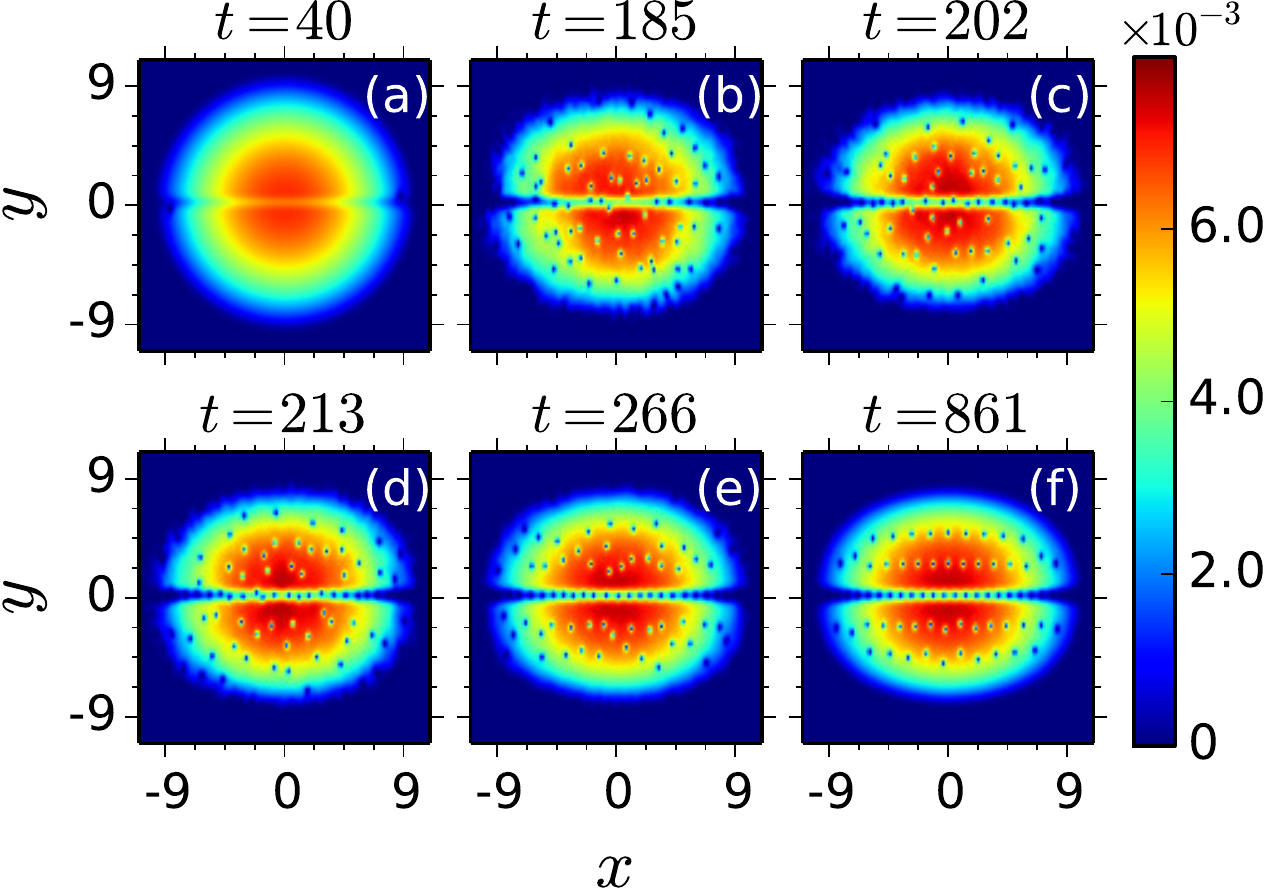}
  \caption{Transport of vortices when laser field energy rate is less
  compared to
  the double well potential depth energy rate with dissipation. Vortices 
  transport between the wells up to the time $t = 266$ ms, after that some 
  vortices are settled near to the interface of the wells. The time (in units 
  of ms) is shown above the plots. Here $x$ and $y$ are measured in units of 
  $a_{\rm osc}$. Density is measured in units of $a_{\rm osc}^{-2}$ and is 
  normalized to unity}
  \label{del_l_g}
  \end{center}
\end{figure}
%
%

\subsubsection{$R_1 < R_2$}
For this case we vary  $U_0$ from $0$ to $25.85$ (in units of $\hbar\omega_x$)
within a period of 202 ms, and the evolution of the density profiles 
are shown in Fig.~(\ref{del_l_g}). The inclusion of the barrier, to form a double 
well potential, accelerates the formation of vortices, and they appear within
a short span of time $\approx 40$ ms. This is much shorter than the time of
$\approx 266$ ms taken to generate vortices in absence of the barrier or in 
harmonic potential as shown in Fig.~(\ref{u0_0_g}). The shortening is due to 
the modified density distribution arising from the presence of the central barrier in the double well potential. The vortices are seeded near the central
region of the barrier where the density is low as shown in 
Fig.~(\ref{del_l_g}) (a). During the quench, at lower values of $U_0$, vortices 
traverse from one well to the other due to the lower depth of the potential, but
it stops once $U_0$ reaches maxima as the vortex energy is not 
enough for transport from one well to the other. The dynamics associated with 
the crystallisation of the  vortices to form a vortex lattice is evident from 
the density patterns in Fig.~(\ref{del_l_g}) (b)-(e). The equilibrium ground 
state solution of vortex lattice is obtained at $\approx 861$ ms after the
free evolution as shown in Fig.~(\ref{del_l_g}) (f). One noticeable feature is 
the confinement of vortices along the barrier with lower spacing compared to the
vortex lattice in the bulk of the condensate. In particular, the spacing between 
vortices is 1.43 $a_{\rm osc}$ and 0.87 $a_{\rm osc}$ in the bulk and in the 
barrier respectively.

\begin{figure}
  \begin{center}
  \includegraphics[width=8.5cm]{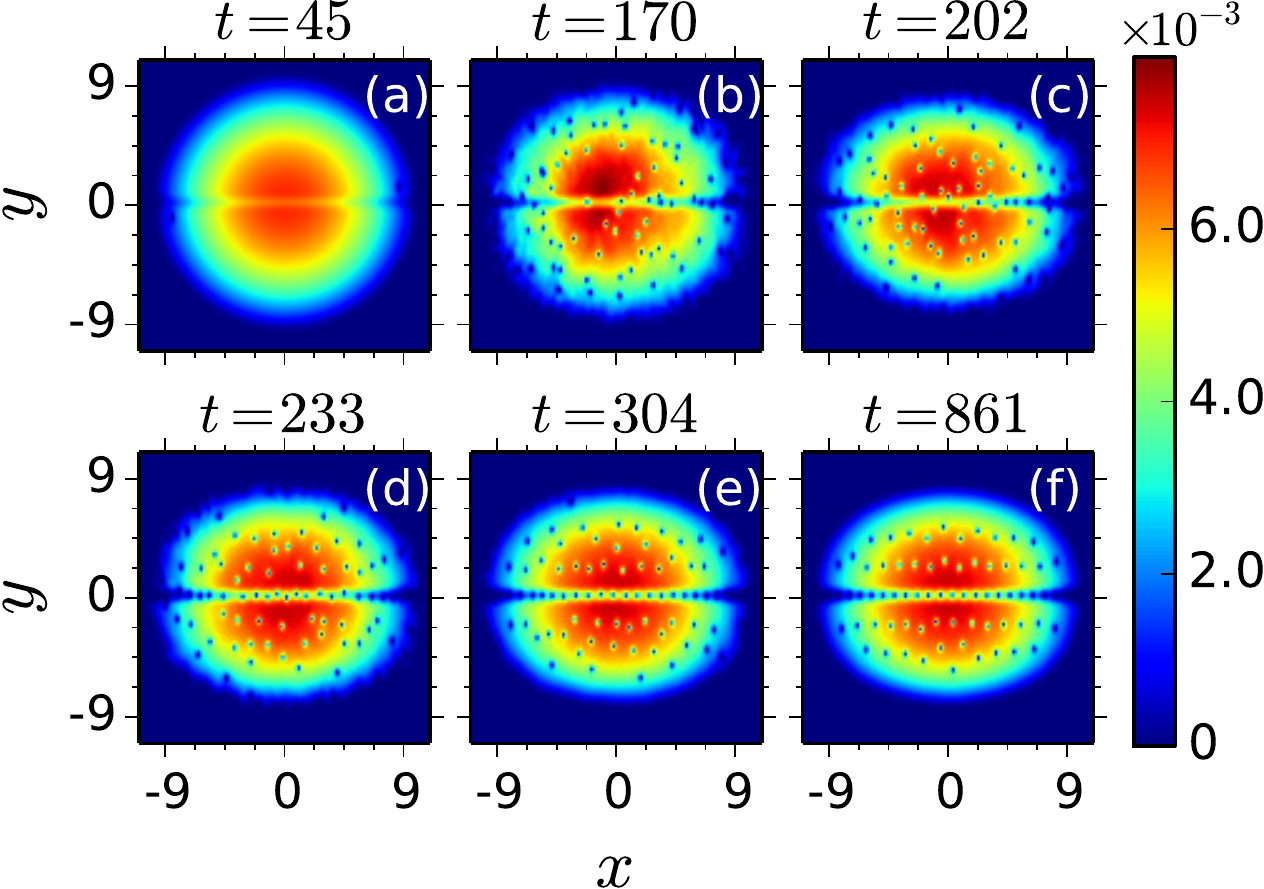}
  \caption{Transport of vortices when laser field energy rate is equal to
  the double well potential depth energy rate with dissipation. Vortices cross 
  from one well to another well up to the time $t = 304$ ms, after that some 
  vortices are settled near the interface. The time (in units of ms) is shown 
  above the plots. Here $x$ and $y$ are measured in units of $a_{\rm osc}$. 
  Density is measured in units of $a_{\rm osc}^{-2}$ and is normalized to 
  unity.}
  \label{del_eq_g}
  \end{center}
\end{figure}
%
\begin{figure}
  \begin{center}
  \includegraphics[width=8.5cm]{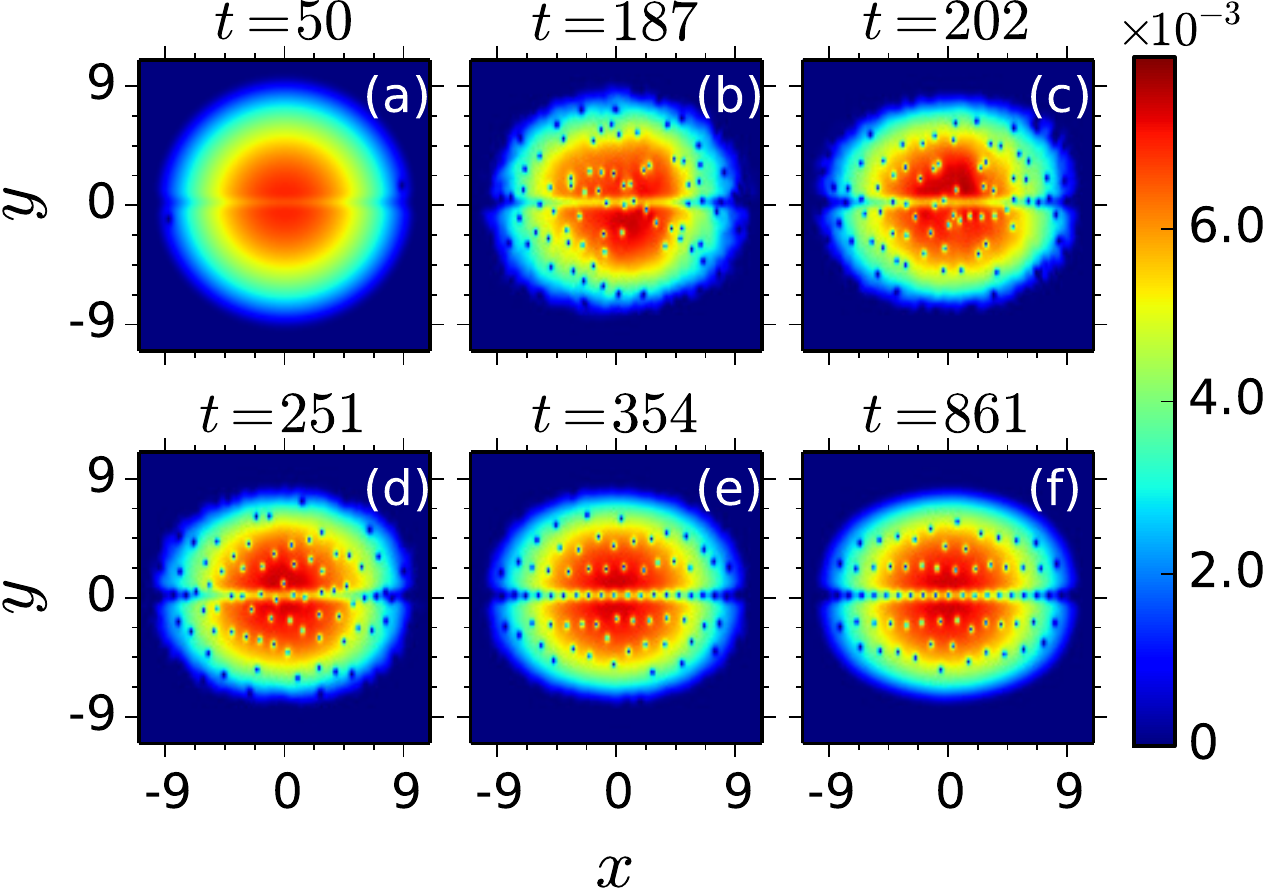}
  \caption{Transport of vortices when laser field energy rate is high
  compared to the double well potential depth energy rate with dissipation. 
  Vortices cross from one well to another well up to the time 
  $t = 354$ ms, after that some vortices are settled near to the interface. 
  The time (in units of ms) is 
  shown above the plots. Here $x$ and $y$ are measured in units of 
  $a_{\rm osc}$.
  Density is measured in units of $a_{\rm osc}^{-2}$ and is normalized to
  unity}
  \label{del_h_g}
  \end{center}
\end{figure}
%

\subsubsection{$R_1 \geqslant R_2$}
For the case of $R_1=R_2$, the value of $U_0$ at the end of the quench is 
$18.85$ (in units of $\hbar\omega_x$). During the quench vortices are generated 
at $\approx 45$ ms of the dynamical evolution, and emerge from 
within the barrier region. Here, the potential depth is less compared to the 
case  of $R_1 < R_2$, and the vortex transportation between the two wells 
occurs for a longer time, that is up to $\approx 304$ ms. Like in the case of 
$R_1 < R_2$ the equilibrium ground state solution is attained 
at $\approx 861$ ms.
For illustration the condensate density profiles during the dynamical evolution
are shown in Fig.~(\ref{del_eq_g}). In the case of $R_1 > R_2$ the value of $U_0$ 
at the end of the quench is $11.85$ (in units of $\hbar\omega_x$). This implies 
that the barrier height attained at the end of the quench is less than the two 
previous cases. The generation of vortices start at $\approx 50$ms, and the 
transportation of vortices between the two wells continues for much longer time, 
till $\approx 354$ ms. Like in the previous cases the equilibrium ground state 
solution is obtained at $\approx 861$ ms as shown in Fig.~(\ref{del_h_g}).


\subsection{Uniform BEC}
\begin{figure}
  \begin{center}
  \includegraphics[width=8.5cm]{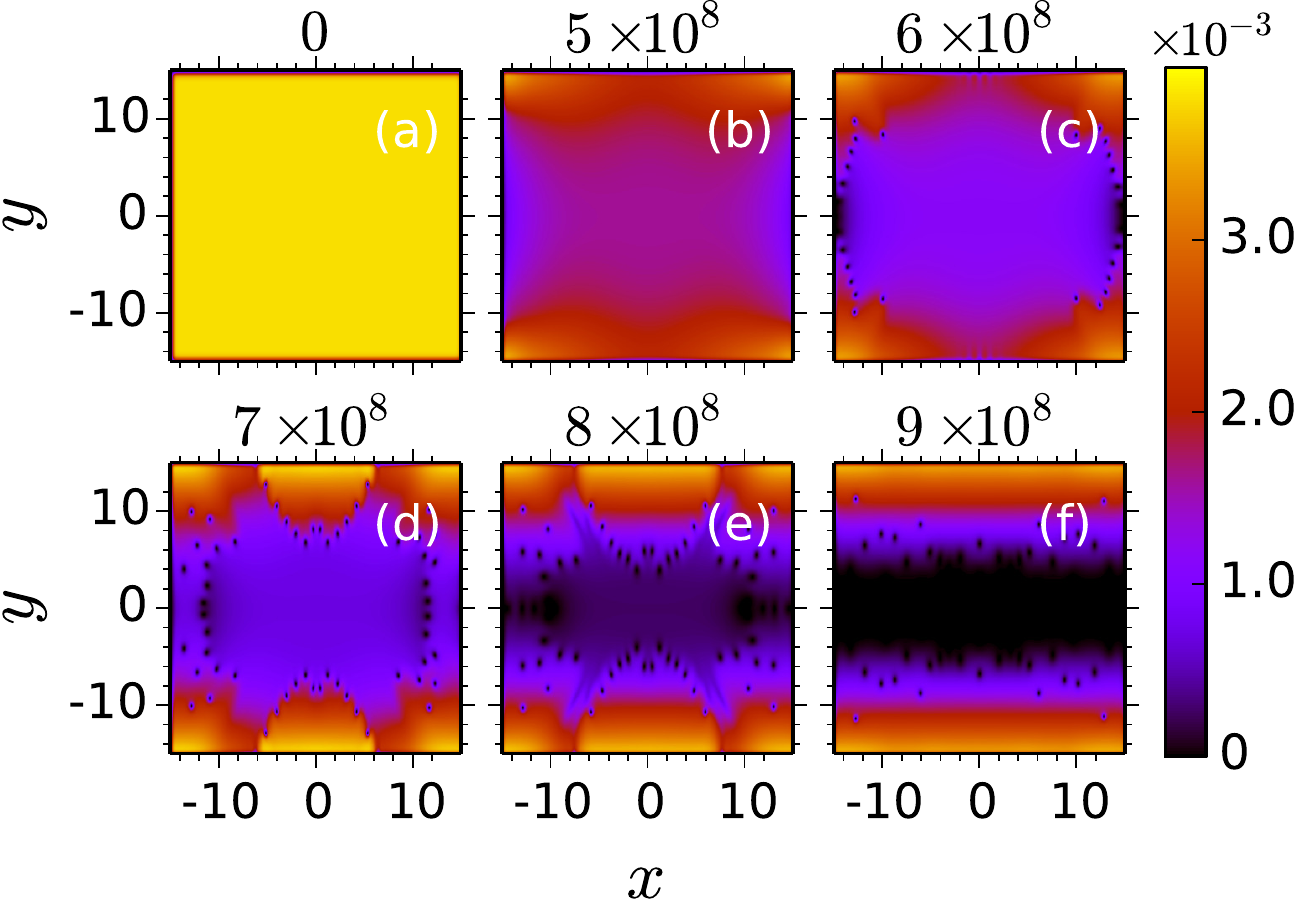}
  \caption{Generation of vortices in the homogeneous system with the hard
          wall boundary with the synthetic magnetic field. The equilibrium 
          solution for different value of $\delta^{'}$(in Hz/m) are shown
          here and values of $\delta^{'}$ are written above each of the plot.
          Vortices are generated near to the boundary and propagate into the
          bulk of the condensate. Here $x$ and $y$ are measured in units of 
          $a_{\rm osc}$. Density is measured in units of 
          $a_{\rm osc}^{-2}$ and is normalized to unity}
  \label{homo}
  \end{center}
\end{figure}

\begin{figure}
  \begin{center}
  \includegraphics[width=8.5cm]{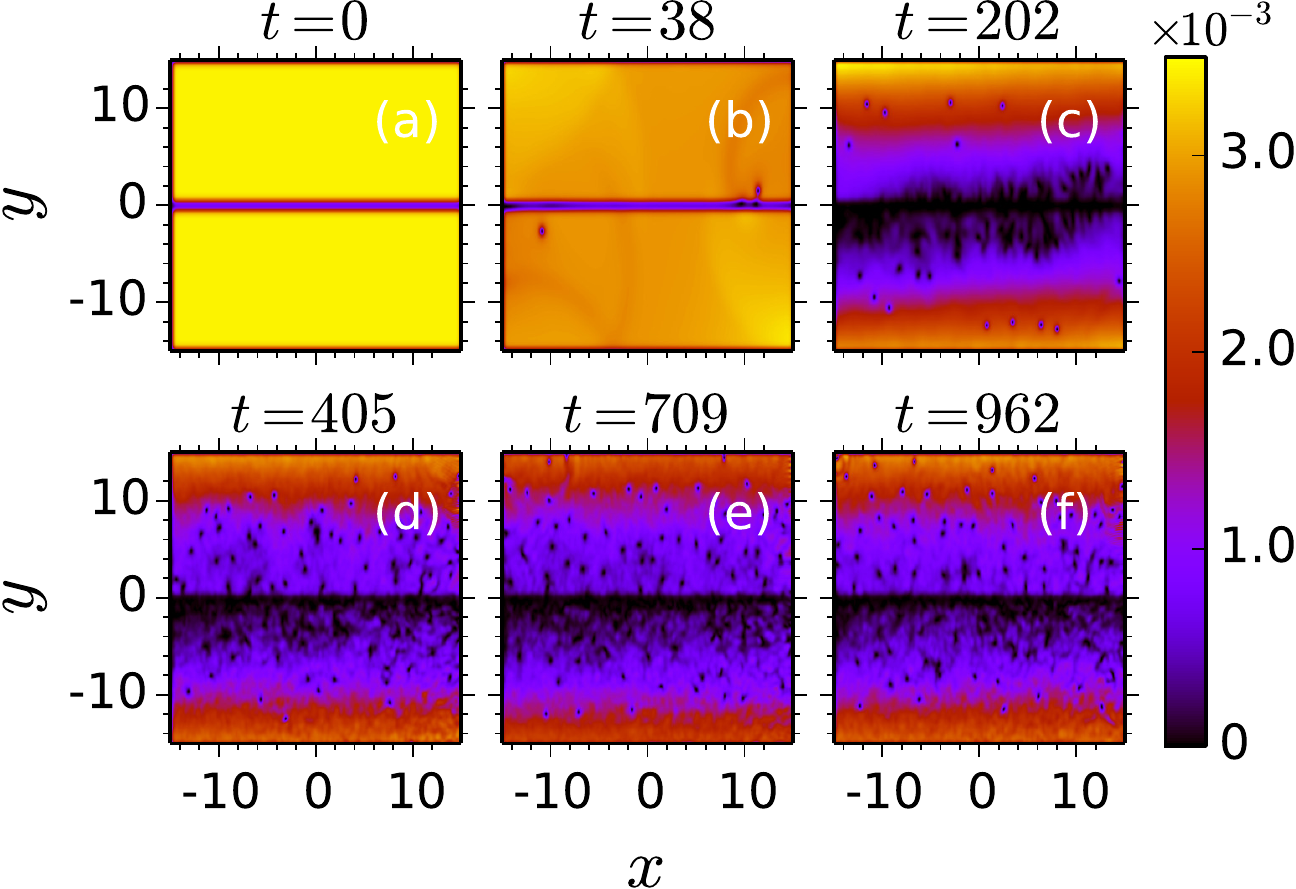}
  \caption{Generation of vortices in the homogeneous system with the hard
          wall boundary. A barrier is introduced along the $x$ direction.
          Vortices are generated near to the barrier and propagate into the
          bulk of the condensate. Here, we do not observe the vortex
          lattice. Here $x$ and $y$ are measured in units of $a_{\rm osc}$.
          Density is measured in units of $a_{\rm osc}^{-2}$ and is 
          normalized to unity.}
  \label{homo_bar}
  \end{center}
\end{figure}

 For uniform BEC, $V_{\rm trap}$ in Eq.~\ref{scale_gp} is set to zero and 
consider hard wall boundary. With this the BEC is uniform except at the 
boundary, where the density goes to zero over the length scale of healing
length. So, the vortices enter from the edges and propagate to the bulk. 
We obtain the equilibrium state of the modified GP Eq.~\ref{scale_gp} with
different values of synthetic magnetic field shown in the Fig.~\ref{homo}. 
We observe that condensate is fragmented at the large value of 
$\delta^{'} > 9\times10^{8}$ Hz/m. In this case there is no formation of vortex 
lattice. We find that the vortices have higher energies in the uniform BEC
$\approx$ 10 $\hbar\omega_x$, whereas it is $\approx$ 3
$\hbar\omega_x$ for BEC with harmonic confining potential. The difference
can accounted by the higher moment of inertia associated with the 
uniform BEC. Next, to compare with the results in presence of harmonic 
potential, we introduce a Gaussian barrier along $x$ axis with 
$U_0 = 10$ (in units of $\hbar \omega_x$) and width of $0.7 \mu $m. 
In the numerical simulation, the initial states at time $t$ = 0 ms is without 
the synthetic magnetic field $\delta^{'}$ = 0 as shown in 
Fig.~\ref{homo_bar}(a). Then, the magnetic field is introduced by quenching 
$\delta^{'}$. The vortices are nucleated at a critical value of $\delta^{'}$ 
as shown in Fig.~\ref{homo_bar}(b) at $t$ = 38 ms. We increase $\delta^{'}$ 
from 0 to $7 \times 10^{8}$ Hz/m in 0 to 202 ms time. After that, we freely 
evolve the system. In the uniform BEC, like in the previous case, the 
vortices nucleate close to the barrier and then propagate to the bulk. 
However, as to be expected, the dynamics of the vortices are qualitatively
different from the inhomogeneous case. The dynamics of the vortices are
determined by the inter-vortex interactions and remain within the bulk regions.
The selected snap shots of the dynamical evolution of the vortices are shown 
in the Fig.~\ref{homo_bar}(a)-(f).


%

\subsection{Critical value of $\delta^{'}$ and density}
As described earlier, we take the equilibrium imaginary time solution of a 
quasi-2D BEC, and evolve it in real time with the introduction of a quenched
artificial gauge potential. This is achieved by increasing the detuning 
gradient of the Raman lasers $\delta '$ and vortices are generated in the 
condensates when $\delta '$ reaches a critical value during the quench. In the
case of a purely harmonic confining potential the critical 
value of $\delta '$ is $0.89\times 10^9$ Hz/m. But, for the case of a condensate
confined in double well potential the critical value of $\delta '$ is  
$0.26\times 10^9$ Hz/m. Hence, the presence of the barrier in the double
well potential lowers the critical value of $\delta '$, which is the 
measure of synthetic magnetic field in the system. To examine the generation of
the vortices in more detail we examine the condensate density at the region
where vortex enters in the condensate. In the case of harmonic trap, vortex
enters from the peripheral region, and we use TF correction to compute the 
density $n_c$. We find that densities $n_c$ for the two trapping 
potentials are $0.1 \times 10^{-3}$ and $0.4 \times 10^{-3}$ for the 
harmonic and double well potential. Here, densities are measured in units of
$a_{\rm osc}^{-2}$ and are normalized to unity. These densities correspond
to the region at which vortex enters in the condensates.\\
%

\section{Conclusions}

 We have shown that the presence of the Gaussian potential barrier enhances the
generation of vortices due to the presence of artificial gauge potential. We
examine this by quenching the artificial gauge potential along with the height
of the barrier potential. Without the barrier potential, in the case of a 
harmonic confining potential, the vortices are generated at a later time 
and vortices are less in number as well. 
Like in the previous works \cite{choi_98,tsubota_02,yan_14}, we observe that 
it is essential to introduce dissipation to obtain equilibrium vortex 
configuration in trapped system. The dissipation drains energy gained 
during the quench and allows the condensate relax to the ground state 
configuration by forming a vortex lattice. In case of uniform BEC, there is 
no formation of vortex lattice even in the presence of dissipation. 
%

\begin{acknowledgements}
We thank Kuldeep Suthar, Sukla Pal and Soumik Bandopadhyay for very useful
discussions. The numerical computations reported in the paper were carried on 
the Vikram-100 HPC cluster at PRL. The work of PM forms a part 
of Science \& Engineering Research Board (SERB), Department of Science \& 
Technology (DST), Govt. of India sponsored research project 
(No. EMR/2014/000644).
\end{acknowledgements}


\bibliography{gauge}{}

\begin{thebibliography}{46}%
\makeatletter
\providecommand \@ifxundefined [1]{%
 \@ifx{#1\undefined}
}%
\providecommand \@ifnum [1]{%
 \ifnum #1\expandafter \@firstoftwo
 \else \expandafter \@secondoftwo
 \fi
}%
\providecommand \@ifx [1]{%
 \ifx #1\expandafter \@firstoftwo
 \else \expandafter \@secondoftwo
 \fi
}%
\providecommand \natexlab [1]{#1}%
\providecommand \enquote  [1]{``#1''}%
\providecommand \bibnamefont  [1]{#1}%
\providecommand \bibfnamefont [1]{#1}%
\providecommand \citenamefont [1]{#1}%
\providecommand \href@noop [0]{\@secondoftwo}%
\providecommand \href [0]{\begingroup \@sanitize@url \@href}%
\providecommand \@href[1]{\@@startlink{#1}\@@href}%
\providecommand \@@href[1]{\endgroup#1\@@endlink}%
\providecommand \@sanitize@url [0]{\catcode `\\12\catcode `\$12\catcode
  `\&12\catcode `\#12\catcode `\^12\catcode `\_12\catcode `\%12\relax}%
\providecommand \@@startlink[1]{}%
\providecommand \@@endlink[0]{}%
\providecommand \url  [0]{\begingroup\@sanitize@url \@url }%
\providecommand \@url [1]{\endgroup\@href {#1}{\urlprefix }}%
\providecommand \urlprefix  [0]{URL }%
\providecommand \Eprint [0]{\href }%
\providecommand \doibase [0]{http://dx.doi.org/}%
\providecommand \selectlanguage [0]{\@gobble}%
\providecommand \bibinfo  [0]{\@secondoftwo}%
\providecommand \bibfield  [0]{\@secondoftwo}%
\providecommand \translation [1]{[#1]}%
\providecommand \BibitemOpen [0]{}%
\providecommand \bibitemStop [0]{}%
\providecommand \bibitemNoStop [0]{.\EOS\space}%
\providecommand \EOS [0]{\spacefactor3000\relax}%
\providecommand \BibitemShut  [1]{\csname bibitem#1\endcsname}%
\let\auto@bib@innerbib\@empty
\bibitem [{\citenamefont {von Klitzing}(1986)}]{klaus_86}%
  \BibitemOpen
  \bibfield  {author} {\bibinfo {author} {\bibfnamefont {K.}~\bibnamefont {von
  Klitzing}},\ }\href {\doibase 10.1103/RevModPhys.58.519} {\bibfield
  {journal} {\bibinfo  {journal} {Rev. Mod. Phys.}\ }\textbf {\bibinfo {volume}
  {58}},\ \bibinfo {pages} {519} (\bibinfo {year} {1986})}\BibitemShut
  {NoStop}%
\bibitem [{\citenamefont {Yennie}(1987)}]{yennie_87}%
  \BibitemOpen
  \bibfield  {author} {\bibinfo {author} {\bibfnamefont {D.~R.}\ \bibnamefont
  {Yennie}},\ }\href {\doibase 10.1103/RevModPhys.59.781} {\bibfield  {journal}
  {\bibinfo  {journal} {Rev. Mod. Phys.}\ }\textbf {\bibinfo {volume} {59}},\
  \bibinfo {pages} {781} (\bibinfo {year} {1987})}\BibitemShut {NoStop}%
\bibitem [{\citenamefont {Stormer}\ \emph {et~al.}(1999)\citenamefont
  {Stormer}, \citenamefont {Tsui},\ and\ \citenamefont
  {Gossard}}]{stormer_99a}%
  \BibitemOpen
  \bibfield  {author} {\bibinfo {author} {\bibfnamefont {H.~L.}\ \bibnamefont
  {Stormer}}, \bibinfo {author} {\bibfnamefont {D.~C.}\ \bibnamefont {Tsui}}, \
  and\ \bibinfo {author} {\bibfnamefont {A.~C.}\ \bibnamefont {Gossard}},\
  }\href {\doibase 10.1103/RevModPhys.71.S298} {\bibfield  {journal} {\bibinfo
  {journal} {Rev. Mod. Phys.}\ }\textbf {\bibinfo {volume} {71}},\ \bibinfo
  {pages} {S298} (\bibinfo {year} {1999})}\BibitemShut {NoStop}%
\bibitem [{\citenamefont {Stormer}(1999)}]{stormer_99b}%
  \BibitemOpen
  \bibfield  {author} {\bibinfo {author} {\bibfnamefont {H.~L.}\ \bibnamefont
  {Stormer}},\ }\href {\doibase 10.1103/RevModPhys.71.875} {\bibfield
  {journal} {\bibinfo  {journal} {Rev. Mod. Phys.}\ }\textbf {\bibinfo {volume}
  {71}},\ \bibinfo {pages} {875} (\bibinfo {year} {1999})}\BibitemShut
  {NoStop}%
\bibitem [{\citenamefont {König}\ \emph {et~al.}(2008)\citenamefont {König},
  \citenamefont {Buhmann}, \citenamefont {Molenkamp}, \citenamefont {Hughes},
  \citenamefont {Liu}, \citenamefont {Qi},\ and\ \citenamefont
  {Zhang}}]{konig_08}%
  \BibitemOpen
  \bibfield  {author} {\bibinfo {author} {\bibfnamefont {M.}~\bibnamefont
  {König}}, \bibinfo {author} {\bibfnamefont {H.}~\bibnamefont {Buhmann}},
  \bibinfo {author} {\bibfnamefont {L.~W.}\ \bibnamefont {Molenkamp}}, \bibinfo
  {author} {\bibfnamefont {T.}~\bibnamefont {Hughes}}, \bibinfo {author}
  {\bibfnamefont {C.-X.}\ \bibnamefont {Liu}}, \bibinfo {author} {\bibfnamefont
  {X.-L.}\ \bibnamefont {Qi}}, \ and\ \bibinfo {author} {\bibfnamefont {S.-C.}\
  \bibnamefont {Zhang}},\ }\href {\doibase 10.1143/JPSJ.77.031007} {\bibfield
  {journal} {\bibinfo  {journal} {J. Phys. Soc. Japan}\ }\textbf {\bibinfo
  {volume} {77}},\ \bibinfo {pages} {031007} (\bibinfo {year}
  {2008})}\BibitemShut {NoStop}%
\bibitem [{\citenamefont {Juzeli\ifmmode~\bar{u}\else \={u}\fi{}nas}\ and\
  \citenamefont {\"Ohberg}(2004)}]{juzeliunas_04}%
  \BibitemOpen
  \bibfield  {author} {\bibinfo {author} {\bibfnamefont {G.}~\bibnamefont
  {Juzeli\ifmmode~\bar{u}\else \={u}\fi{}nas}}\ and\ \bibinfo {author}
  {\bibfnamefont {P.}~\bibnamefont {\"Ohberg}},\ }\href {\doibase
  10.1103/PhysRevLett.93.033602} {\bibfield  {journal} {\bibinfo  {journal}
  {Phys. Rev. Lett.}\ }\textbf {\bibinfo {volume} {93}},\ \bibinfo {pages}
  {033602} (\bibinfo {year} {2004})}\BibitemShut {NoStop}%
\bibitem [{\citenamefont {Zhu}\ \emph {et~al.}(2006)\citenamefont {Zhu},
  \citenamefont {Fu}, \citenamefont {Wu}, \citenamefont {Zhang},\ and\
  \citenamefont {Duan}}]{zhu_06}%
  \BibitemOpen
  \bibfield  {author} {\bibinfo {author} {\bibfnamefont {S.-L.}\ \bibnamefont
  {Zhu}}, \bibinfo {author} {\bibfnamefont {H.}~\bibnamefont {Fu}}, \bibinfo
  {author} {\bibfnamefont {C.-J.}\ \bibnamefont {Wu}}, \bibinfo {author}
  {\bibfnamefont {S.-C.}\ \bibnamefont {Zhang}}, \ and\ \bibinfo {author}
  {\bibfnamefont {L.-M.}\ \bibnamefont {Duan}},\ }\href {\doibase
  10.1103/PhysRevLett.97.240401} {\bibfield  {journal} {\bibinfo  {journal}
  {Phys. Rev. Lett.}\ }\textbf {\bibinfo {volume} {97}},\ \bibinfo {pages}
  {240401} (\bibinfo {year} {2006})}\BibitemShut {NoStop}%
\bibitem [{\citenamefont {Spielman}(2009)}]{spielman_09}%
  \BibitemOpen
  \bibfield  {author} {\bibinfo {author} {\bibfnamefont {I.~B.}\ \bibnamefont
  {Spielman}},\ }\href {\doibase 10.1103/PhysRevA.79.063613} {\bibfield
  {journal} {\bibinfo  {journal} {Phys. Rev. A}\ }\textbf {\bibinfo {volume}
  {79}},\ \bibinfo {pages} {063613} (\bibinfo {year} {2009})}\BibitemShut
  {NoStop}%
\bibitem [{\citenamefont {Jim´enez-Garc´ıa}(2012)}]{karina_12}%
  \BibitemOpen
  \bibfield  {author} {\bibinfo {author} {\bibfnamefont {K.}~\bibnamefont
  {Jim´enez-Garc´ıa}},\ }\emph {\bibinfo {title} {{Artificial Gauge Fields
  for Ultracold Neutral Atoms}}},\ \href@noop {} {Ph.D. thesis},\ \bibinfo
  {school} {National Institute of Standards and Technology, and the University
  of Maryland Gaithersburg, Maryland} (\bibinfo {year} {2012})\BibitemShut
  {NoStop}%
\bibitem [{\citenamefont {Goldman}\ \emph {et~al.}(2014)\citenamefont
  {Goldman}, \citenamefont {Juzeliūnas}, \citenamefont {{\"O}hberg},\ and\
  \citenamefont {Spielman}}]{goldman_14}%
  \BibitemOpen
  \bibfield  {author} {\bibinfo {author} {\bibfnamefont {N.}~\bibnamefont
  {Goldman}}, \bibinfo {author} {\bibfnamefont {G.}~\bibnamefont
  {Juzeliūnas}}, \bibinfo {author} {\bibfnamefont {P.}~\bibnamefont
  {{\"O}hberg}}, \ and\ \bibinfo {author} {\bibfnamefont {I.~B.}\ \bibnamefont
  {Spielman}},\ }\href@noop {} {\bibfield  {journal} {\bibinfo  {journal} {Rep.
  Prog. Phys.}\ }\textbf {\bibinfo {volume} {77}},\ \bibinfo {pages} {126401}
  (\bibinfo {year} {2014})}\BibitemShut {NoStop}%
\bibitem [{\citenamefont {Madison}\ \emph {et~al.}(2000)\citenamefont
  {Madison}, \citenamefont {Chevy}, \citenamefont {Wohlleben},\ and\
  \citenamefont {Dalibard}}]{madison_00}%
  \BibitemOpen
  \bibfield  {author} {\bibinfo {author} {\bibfnamefont {K.~W.}\ \bibnamefont
  {Madison}}, \bibinfo {author} {\bibfnamefont {F.}~\bibnamefont {Chevy}},
  \bibinfo {author} {\bibfnamefont {W.}~\bibnamefont {Wohlleben}}, \ and\
  \bibinfo {author} {\bibfnamefont {J.}~\bibnamefont {Dalibard}},\ }\href
  {\doibase 10.1103/PhysRevLett.84.806} {\bibfield  {journal} {\bibinfo
  {journal} {Phys. Rev. Lett.}\ }\textbf {\bibinfo {volume} {84}},\ \bibinfo
  {pages} {806} (\bibinfo {year} {2000})}\BibitemShut {NoStop}%
\bibitem [{\citenamefont {Abo-Shaeer}\ \emph {et~al.}(2001)\citenamefont
  {Abo-Shaeer}, \citenamefont {Raman}, \citenamefont {Vogels},\ and\
  \citenamefont {Ketterle}}]{shaeer_01}%
  \BibitemOpen
  \bibfield  {author} {\bibinfo {author} {\bibfnamefont {J.~R.}\ \bibnamefont
  {Abo-Shaeer}}, \bibinfo {author} {\bibfnamefont {C.}~\bibnamefont {Raman}},
  \bibinfo {author} {\bibfnamefont {J.~M.}\ \bibnamefont {Vogels}}, \ and\
  \bibinfo {author} {\bibfnamefont {W.}~\bibnamefont {Ketterle}},\ }\href
  {\doibase 10.1126/science.1060182} {\bibfield  {journal} {\bibinfo  {journal}
  {Science}\ }\textbf {\bibinfo {volume} {292}},\ \bibinfo {pages} {476}
  (\bibinfo {year} {2001})}\BibitemShut {NoStop}%
\bibitem [{\citenamefont {Haljan}\ \emph {et~al.}(2001)\citenamefont {Haljan},
  \citenamefont {Coddington}, \citenamefont {Engels},\ and\ \citenamefont
  {Cornell}}]{haljan_01}%
  \BibitemOpen
  \bibfield  {author} {\bibinfo {author} {\bibfnamefont {P.~C.}\ \bibnamefont
  {Haljan}}, \bibinfo {author} {\bibfnamefont {I.}~\bibnamefont {Coddington}},
  \bibinfo {author} {\bibfnamefont {P.}~\bibnamefont {Engels}}, \ and\ \bibinfo
  {author} {\bibfnamefont {E.~A.}\ \bibnamefont {Cornell}},\ }\href {\doibase
  10.1103/PhysRevLett.87.210403} {\bibfield  {journal} {\bibinfo  {journal}
  {Phys. Rev. Lett.}\ }\textbf {\bibinfo {volume} {87}},\ \bibinfo {pages}
  {210403} (\bibinfo {year} {2001})}\BibitemShut {NoStop}%
\bibitem [{\citenamefont {Isoshima}\ \emph {et~al.}(2000)\citenamefont
  {Isoshima}, \citenamefont {Nakahara}, \citenamefont {Ohmi},\ and\
  \citenamefont {Machida}}]{isoshima_00}%
  \BibitemOpen
  \bibfield  {author} {\bibinfo {author} {\bibfnamefont {T.}~\bibnamefont
  {Isoshima}}, \bibinfo {author} {\bibfnamefont {M.}~\bibnamefont {Nakahara}},
  \bibinfo {author} {\bibfnamefont {T.}~\bibnamefont {Ohmi}}, \ and\ \bibinfo
  {author} {\bibfnamefont {K.}~\bibnamefont {Machida}},\ }\href {\doibase
  10.1103/PhysRevA.61.063610} {\bibfield  {journal} {\bibinfo  {journal} {Phys.
  Rev. A}\ }\textbf {\bibinfo {volume} {61}},\ \bibinfo {pages} {063610}
  (\bibinfo {year} {2000})}\BibitemShut {NoStop}%
\bibitem [{\citenamefont {Leanhardt}\ \emph {et~al.}(2002)\citenamefont
  {Leanhardt}, \citenamefont {G\"orlitz}, \citenamefont {Chikkatur},
  \citenamefont {Kielpinski}, \citenamefont {Shin}, \citenamefont {Pritchard},\
  and\ \citenamefont {Ketterle}}]{lea_02}%
  \BibitemOpen
  \bibfield  {author} {\bibinfo {author} {\bibfnamefont {A.~E.}\ \bibnamefont
  {Leanhardt}}, \bibinfo {author} {\bibfnamefont {A.}~\bibnamefont
  {G\"orlitz}}, \bibinfo {author} {\bibfnamefont {A.~P.}\ \bibnamefont
  {Chikkatur}}, \bibinfo {author} {\bibfnamefont {D.}~\bibnamefont
  {Kielpinski}}, \bibinfo {author} {\bibfnamefont {Y.}~\bibnamefont {Shin}},
  \bibinfo {author} {\bibfnamefont {D.~E.}\ \bibnamefont {Pritchard}}, \ and\
  \bibinfo {author} {\bibfnamefont {W.}~\bibnamefont {Ketterle}},\ }\href
  {\doibase 10.1103/PhysRevLett.89.190403} {\bibfield  {journal} {\bibinfo
  {journal} {Phys. Rev. Lett.}\ }\textbf {\bibinfo {volume} {89}},\ \bibinfo
  {pages} {190403} (\bibinfo {year} {2002})}\BibitemShut {NoStop}%
\bibitem [{\citenamefont {Williams}\ and\ \citenamefont
  {Holland}(1999)}]{williams_99}%
  \BibitemOpen
  \bibfield  {author} {\bibinfo {author} {\bibfnamefont {J.~E.}\ \bibnamefont
  {Williams}}\ and\ \bibinfo {author} {\bibfnamefont {M.~J.}\ \bibnamefont
  {Holland}},\ }\href {http://dx.doi.org/10.1038/44095} {\bibfield  {journal}
  {\bibinfo  {journal} {Nature}\ }\textbf {\bibinfo {volume} {401}},\ \bibinfo
  {pages} {568} (\bibinfo {year} {1999})}\BibitemShut {NoStop}%
\bibitem [{\citenamefont {Matthews}\ \emph {et~al.}(1999)\citenamefont
  {Matthews}, \citenamefont {Anderson}, \citenamefont {Haljan}, \citenamefont
  {Hall}, \citenamefont {Wieman},\ and\ \citenamefont {Cornell}}]{matthews_99}%
  \BibitemOpen
  \bibfield  {author} {\bibinfo {author} {\bibfnamefont {M.~R.}\ \bibnamefont
  {Matthews}}, \bibinfo {author} {\bibfnamefont {B.~P.}\ \bibnamefont
  {Anderson}}, \bibinfo {author} {\bibfnamefont {P.~C.}\ \bibnamefont
  {Haljan}}, \bibinfo {author} {\bibfnamefont {D.~S.}\ \bibnamefont {Hall}},
  \bibinfo {author} {\bibfnamefont {C.~E.}\ \bibnamefont {Wieman}}, \ and\
  \bibinfo {author} {\bibfnamefont {E.~A.}\ \bibnamefont {Cornell}},\ }\href
  {\doibase 10.1103/PhysRevLett.83.2498} {\bibfield  {journal} {\bibinfo
  {journal} {Phys. Rev. Lett.}\ }\textbf {\bibinfo {volume} {83}},\ \bibinfo
  {pages} {2498} (\bibinfo {year} {1999})}\BibitemShut {NoStop}%
\bibitem [{\citenamefont {Price}\ \emph {et~al.}(2016)\citenamefont {Price},
  \citenamefont {Trypogeorgos}, \citenamefont {Campbell}, \citenamefont
  {Putra}, \citenamefont {Valdés-Curiel},\ and\ \citenamefont
  {Spielman}}]{price_16}%
  \BibitemOpen
  \bibfield  {author} {\bibinfo {author} {\bibfnamefont {R.~M.}\ \bibnamefont
  {Price}}, \bibinfo {author} {\bibfnamefont {D.}~\bibnamefont {Trypogeorgos}},
  \bibinfo {author} {\bibfnamefont {D.~L.}\ \bibnamefont {Campbell}}, \bibinfo
  {author} {\bibfnamefont {A.}~\bibnamefont {Putra}}, \bibinfo {author}
  {\bibfnamefont {A.}~\bibnamefont {Valdés-Curiel}}, \ and\ \bibinfo {author}
  {\bibfnamefont {I.~B.}\ \bibnamefont {Spielman}},\ }\href
  {http://stacks.iop.org/1367-2630/18/i=11/a=113009} {\bibfield  {journal}
  {\bibinfo  {journal} {New J. Phys.}\ }\textbf {\bibinfo {volume} {18}},\
  \bibinfo {pages} {113009} (\bibinfo {year} {2016})}\BibitemShut {NoStop}%
\bibitem [{\citenamefont {Tsubota}\ \emph {et~al.}(2013)\citenamefont
  {Tsubota}, \citenamefont {Kobayashi},\ and\ \citenamefont
  {Takeuchi}}]{tsubota_13}%
  \BibitemOpen
  \bibfield  {author} {\bibinfo {author} {\bibfnamefont {M.}~\bibnamefont
  {Tsubota}}, \bibinfo {author} {\bibfnamefont {M.}~\bibnamefont {Kobayashi}},
  \ and\ \bibinfo {author} {\bibfnamefont {H.}~\bibnamefont {Takeuchi}},\
  }\href {\doibase https://doi.org/10.1016/j.physrep.2012.09.007} {\bibfield
  {journal} {\bibinfo  {journal} {Physics Reports}\ }\textbf {\bibinfo {volume}
  {522}},\ \bibinfo {pages} {191 } (\bibinfo {year} {2013})}\BibitemShut
  {NoStop}%
\bibitem [{\citenamefont {Nemirovskii}(2013)}]{nemirovskii_13}%
  \BibitemOpen
  \bibfield  {author} {\bibinfo {author} {\bibfnamefont {S.~K.}\ \bibnamefont
  {Nemirovskii}},\ }\href {\doibase
  https://doi.org/10.1016/j.physrep.2012.10.005} {\bibfield  {journal}
  {\bibinfo  {journal} {Physics Reports}\ }\textbf {\bibinfo {volume} {524}},\
  \bibinfo {pages} {85 } (\bibinfo {year} {2013})}\BibitemShut {NoStop}%
\bibitem [{\citenamefont {Javanainen}(1986)}]{juha_86}%
  \BibitemOpen
  \bibfield  {author} {\bibinfo {author} {\bibfnamefont {J.}~\bibnamefont
  {Javanainen}},\ }\href {\doibase 10.1103/PhysRevLett.57.3164} {\bibfield
  {journal} {\bibinfo  {journal} {Phys. Rev. Lett.}\ }\textbf {\bibinfo
  {volume} {57}},\ \bibinfo {pages} {3164} (\bibinfo {year}
  {1986})}\BibitemShut {NoStop}%
\bibitem [{\citenamefont {Dalfovo}\ \emph {et~al.}(1996)\citenamefont
  {Dalfovo}, \citenamefont {Pitaevskii},\ and\ \citenamefont
  {Stringari}}]{dalfovo_96}%
  \BibitemOpen
  \bibfield  {author} {\bibinfo {author} {\bibfnamefont {F.}~\bibnamefont
  {Dalfovo}}, \bibinfo {author} {\bibfnamefont {L.}~\bibnamefont {Pitaevskii}},
  \ and\ \bibinfo {author} {\bibfnamefont {S.}~\bibnamefont {Stringari}},\
  }\href {\doibase 10.1103/PhysRevA.54.4213} {\bibfield  {journal} {\bibinfo
  {journal} {Phys. Rev. A}\ }\textbf {\bibinfo {volume} {54}},\ \bibinfo
  {pages} {4213} (\bibinfo {year} {1996})}\BibitemShut {NoStop}%
\bibitem [{\citenamefont {Smerzi}\ \emph {et~al.}(1997)\citenamefont {Smerzi},
  \citenamefont {Fantoni}, \citenamefont {Giovanazzi},\ and\ \citenamefont
  {Shenoy}}]{smerzi_97}%
  \BibitemOpen
  \bibfield  {author} {\bibinfo {author} {\bibfnamefont {A.}~\bibnamefont
  {Smerzi}}, \bibinfo {author} {\bibfnamefont {S.}~\bibnamefont {Fantoni}},
  \bibinfo {author} {\bibfnamefont {S.}~\bibnamefont {Giovanazzi}}, \ and\
  \bibinfo {author} {\bibfnamefont {S.~R.}\ \bibnamefont {Shenoy}},\ }\href
  {\doibase 10.1103/PhysRevLett.79.4950} {\bibfield  {journal} {\bibinfo
  {journal} {Phys. Rev. Lett.}\ }\textbf {\bibinfo {volume} {79}},\ \bibinfo
  {pages} {4950} (\bibinfo {year} {1997})}\BibitemShut {NoStop}%
\bibitem [{\citenamefont {Cataliotti}\ \emph {et~al.}(2001)\citenamefont
  {Cataliotti}, \citenamefont {Burger}, \citenamefont {Fort}, \citenamefont
  {Maddaloni}, \citenamefont {Minardi}, \citenamefont {Trombettoni},
  \citenamefont {Smerzi},\ and\ \citenamefont {Inguscio}}]{cataliotti_01}%
  \BibitemOpen
  \bibfield  {author} {\bibinfo {author} {\bibfnamefont {F.~S.}\ \bibnamefont
  {Cataliotti}}, \bibinfo {author} {\bibfnamefont {S.}~\bibnamefont {Burger}},
  \bibinfo {author} {\bibfnamefont {C.}~\bibnamefont {Fort}}, \bibinfo {author}
  {\bibfnamefont {P.}~\bibnamefont {Maddaloni}}, \bibinfo {author}
  {\bibfnamefont {F.}~\bibnamefont {Minardi}}, \bibinfo {author} {\bibfnamefont
  {A.}~\bibnamefont {Trombettoni}}, \bibinfo {author} {\bibfnamefont
  {A.}~\bibnamefont {Smerzi}}, \ and\ \bibinfo {author} {\bibfnamefont
  {M.}~\bibnamefont {Inguscio}},\ }\href {\doibase 10.1126/science.1062612}
  {\bibfield  {journal} {\bibinfo  {journal} {Science}\ }\textbf {\bibinfo
  {volume} {293}},\ \bibinfo {pages} {843} (\bibinfo {year}
  {2001})}\BibitemShut {NoStop}%
\bibitem [{\citenamefont {Albiez}\ \emph {et~al.}(2005)\citenamefont {Albiez},
  \citenamefont {Gati}, \citenamefont {F\"olling}, \citenamefont {Hunsmann},
  \citenamefont {Cristiani},\ and\ \citenamefont {Oberthaler}}]{albiez_05}%
  \BibitemOpen
  \bibfield  {author} {\bibinfo {author} {\bibfnamefont {M.}~\bibnamefont
  {Albiez}}, \bibinfo {author} {\bibfnamefont {R.}~\bibnamefont {Gati}},
  \bibinfo {author} {\bibfnamefont {J.}~\bibnamefont {F\"olling}}, \bibinfo
  {author} {\bibfnamefont {S.}~\bibnamefont {Hunsmann}}, \bibinfo {author}
  {\bibfnamefont {M.}~\bibnamefont {Cristiani}}, \ and\ \bibinfo {author}
  {\bibfnamefont {M.~K.}\ \bibnamefont {Oberthaler}},\ }\href {\doibase
  10.1103/PhysRevLett.95.010402} {\bibfield  {journal} {\bibinfo  {journal}
  {Phys. Rev. Lett.}\ }\textbf {\bibinfo {volume} {95}},\ \bibinfo {pages}
  {010402} (\bibinfo {year} {2005})}\BibitemShut {NoStop}%
\bibitem [{\citenamefont {Levy}\ \emph {et~al.}(2007)\citenamefont {Levy},
  \citenamefont {Lahoud}, \citenamefont {Shomroni},\ and\ \citenamefont
  {Steinhauer}}]{levy_07}%
  \BibitemOpen
  \bibfield  {author} {\bibinfo {author} {\bibfnamefont {S.}~\bibnamefont
  {Levy}}, \bibinfo {author} {\bibfnamefont {E.}~\bibnamefont {Lahoud}},
  \bibinfo {author} {\bibfnamefont {I.}~\bibnamefont {Shomroni}}, \ and\
  \bibinfo {author} {\bibfnamefont {J.}~\bibnamefont {Steinhauer}},\ }\href
  {\doibase 10.1038/nature06186} {\bibfield  {journal} {\bibinfo  {journal}
  {Nature}\ }\textbf {\bibinfo {volume} {449}},\ \bibinfo {pages} {579}
  (\bibinfo {year} {2007})}\BibitemShut {NoStop}%
\bibitem [{\citenamefont {Trenkwalder}\ \emph {et~al.}(2016)\citenamefont
  {Trenkwalder}, \citenamefont {Spagnolli}, \citenamefont {Semeghini},
  \citenamefont {Coop}, \citenamefont {Landini}, \citenamefont {Castilho},
  \citenamefont {Pezze}, \citenamefont {Modugno}, \citenamefont {Inguscio},
  \citenamefont {Smerzi},\ and\ \citenamefont {Fattori}}]{trenkwalder_16}%
  \BibitemOpen
  \bibfield  {author} {\bibinfo {author} {\bibfnamefont {A.}~\bibnamefont
  {Trenkwalder}}, \bibinfo {author} {\bibfnamefont {G.}~\bibnamefont
  {Spagnolli}}, \bibinfo {author} {\bibfnamefont {G.}~\bibnamefont
  {Semeghini}}, \bibinfo {author} {\bibfnamefont {S.}~\bibnamefont {Coop}},
  \bibinfo {author} {\bibfnamefont {M.}~\bibnamefont {Landini}}, \bibinfo
  {author} {\bibfnamefont {P.}~\bibnamefont {Castilho}}, \bibinfo {author}
  {\bibfnamefont {L.}~\bibnamefont {Pezze}}, \bibinfo {author} {\bibfnamefont
  {G.}~\bibnamefont {Modugno}}, \bibinfo {author} {\bibfnamefont
  {M.}~\bibnamefont {Inguscio}}, \bibinfo {author} {\bibfnamefont
  {A.}~\bibnamefont {Smerzi}}, \ and\ \bibinfo {author} {\bibfnamefont
  {M.}~\bibnamefont {Fattori}},\ }\href {\doibase 10.1038/nphys3743} {\bibfield
   {journal} {\bibinfo  {journal} {Nat. Phys.}\ }\textbf {\bibinfo {volume}
  {12}},\ \bibinfo {pages} {826} (\bibinfo {year} {2016})}\BibitemShut
  {NoStop}%
\bibitem [{\citenamefont {Garcia-March}\ and\ \citenamefont
  {Carr}(2015)}]{garcia}%
  \BibitemOpen
  \bibfield  {author} {\bibinfo {author} {\bibfnamefont {M.~A.}\ \bibnamefont
  {Garcia-March}}\ and\ \bibinfo {author} {\bibfnamefont {L.~D.}\ \bibnamefont
  {Carr}},\ }\href {\doibase 10.1103/PhysRevA.91.033626} {\bibfield  {journal}
  {\bibinfo  {journal} {Phys. Rev. A}\ }\textbf {\bibinfo {volume} {91}},\
  \bibinfo {pages} {033626} (\bibinfo {year} {2015})}\BibitemShut {NoStop}%
\bibitem [{\citenamefont {Andrews}\ \emph {et~al.}(1997)\citenamefont
  {Andrews}, \citenamefont {Townsend}, \citenamefont {Miesner}, \citenamefont
  {Durfee}, \citenamefont {Kurn},\ and\ \citenamefont {Ketterle}}]{andrews_97}%
  \BibitemOpen
  \bibfield  {author} {\bibinfo {author} {\bibfnamefont {M.~R.}\ \bibnamefont
  {Andrews}}, \bibinfo {author} {\bibfnamefont {C.~G.}\ \bibnamefont
  {Townsend}}, \bibinfo {author} {\bibfnamefont {H.-J.}\ \bibnamefont
  {Miesner}}, \bibinfo {author} {\bibfnamefont {D.~S.}\ \bibnamefont {Durfee}},
  \bibinfo {author} {\bibfnamefont {D.~M.}\ \bibnamefont {Kurn}}, \ and\
  \bibinfo {author} {\bibfnamefont {W.}~\bibnamefont {Ketterle}},\ }\href
  {\doibase 10.1126/science.275.5300.637} {\bibfield  {journal} {\bibinfo
  {journal} {Science}\ }\textbf {\bibinfo {volume} {275}},\ \bibinfo {pages}
  {637} (\bibinfo {year} {1997})}\BibitemShut {NoStop}%
\bibitem [{\citenamefont {Nguyen}\ \emph {et~al.}(2014)\citenamefont {Nguyen},
  \citenamefont {Dyke}, \citenamefont {Luo}, \citenamefont {Malomed},\ and\
  \citenamefont {Hulet}}]{nguyen_14}%
  \BibitemOpen
  \bibfield  {author} {\bibinfo {author} {\bibfnamefont {J.~H.~V.}\
  \bibnamefont {Nguyen}}, \bibinfo {author} {\bibfnamefont {P.}~\bibnamefont
  {Dyke}}, \bibinfo {author} {\bibfnamefont {D.}~\bibnamefont {Luo}}, \bibinfo
  {author} {\bibfnamefont {B.~A.}\ \bibnamefont {Malomed}}, \ and\ \bibinfo
  {author} {\bibfnamefont {R.~G.}\ \bibnamefont {Hulet}},\ }\href {\doibase
  10.1038/nphys3135} {\bibfield  {journal} {\bibinfo  {journal} {Nat Phys}\
  }\textbf {\bibinfo {volume} {10}},\ \bibinfo {pages} {918} (\bibinfo {year}
  {2014})}\BibitemShut {NoStop}%
\bibitem [{\citenamefont {Gaunt}\ \emph {et~al.}(2013)\citenamefont {Gaunt},
  \citenamefont {Schmidutz}, \citenamefont {Gotlibovych}, \citenamefont
  {Smith},\ and\ \citenamefont {Hadzibabic}}]{Gaunt_13}%
  \BibitemOpen
  \bibfield  {author} {\bibinfo {author} {\bibfnamefont {A.~L.}\ \bibnamefont
  {Gaunt}}, \bibinfo {author} {\bibfnamefont {T.~F.}\ \bibnamefont
  {Schmidutz}}, \bibinfo {author} {\bibfnamefont {I.}~\bibnamefont
  {Gotlibovych}}, \bibinfo {author} {\bibfnamefont {R.~P.}\ \bibnamefont
  {Smith}}, \ and\ \bibinfo {author} {\bibfnamefont {Z.}~\bibnamefont
  {Hadzibabic}},\ }\href {\doibase 10.1103/PhysRevLett.110.200406} {\bibfield
  {journal} {\bibinfo  {journal} {Phys. Rev. Lett.}\ }\textbf {\bibinfo
  {volume} {110}},\ \bibinfo {pages} {200406} (\bibinfo {year}
  {2013})}\BibitemShut {NoStop}%
\bibitem [{\citenamefont {Navon}\ \emph {et~al.}(2016)\citenamefont {Navon},
  \citenamefont {Gaunt}, \citenamefont {Smith},\ and\ \citenamefont
  {Hadzibabic}}]{navon_16}%
  \BibitemOpen
  \bibfield  {author} {\bibinfo {author} {\bibfnamefont {N.}~\bibnamefont
  {Navon}}, \bibinfo {author} {\bibfnamefont {A.~L.}\ \bibnamefont {Gaunt}},
  \bibinfo {author} {\bibfnamefont {R.~P.}\ \bibnamefont {Smith}}, \ and\
  \bibinfo {author} {\bibfnamefont {Z.}~\bibnamefont {Hadzibabic}},\ }\href
  {\doibase 10.1038/nature20114} {\bibfield  {journal} {\bibinfo  {journal}
  {Nature}\ }\textbf {\bibinfo {volume} {539}},\ \bibinfo {pages} {72}
  (\bibinfo {year} {2016})}\BibitemShut {NoStop}%
\bibitem [{\citenamefont {Choi}\ \emph {et~al.}(1998)\citenamefont {Choi},
  \citenamefont {Morgan},\ and\ \citenamefont {Burnett}}]{choi_98}%
  \BibitemOpen
  \bibfield  {author} {\bibinfo {author} {\bibfnamefont {S.}~\bibnamefont
  {Choi}}, \bibinfo {author} {\bibfnamefont {S.~A.}\ \bibnamefont {Morgan}}, \
  and\ \bibinfo {author} {\bibfnamefont {K.}~\bibnamefont {Burnett}},\ }\href
  {\doibase 10.1103/PhysRevA.57.4057} {\bibfield  {journal} {\bibinfo
  {journal} {Phys. Rev. A}\ }\textbf {\bibinfo {volume} {57}},\ \bibinfo
  {pages} {4057} (\bibinfo {year} {1998})}\BibitemShut {NoStop}%
\bibitem [{\citenamefont {Tsubota}\ \emph {et~al.}(2002)\citenamefont
  {Tsubota}, \citenamefont {Kasamatsu},\ and\ \citenamefont
  {Ueda}}]{tsubota_02}%
  \BibitemOpen
  \bibfield  {author} {\bibinfo {author} {\bibfnamefont {M.}~\bibnamefont
  {Tsubota}}, \bibinfo {author} {\bibfnamefont {K.}~\bibnamefont {Kasamatsu}},
  \ and\ \bibinfo {author} {\bibfnamefont {M.}~\bibnamefont {Ueda}},\ }\href
  {\doibase 10.1103/PhysRevA.65.023603} {\bibfield  {journal} {\bibinfo
  {journal} {Phys. Rev. A}\ }\textbf {\bibinfo {volume} {65}},\ \bibinfo
  {pages} {023603} (\bibinfo {year} {2002})}\BibitemShut {NoStop}%
\bibitem [{\citenamefont {Yan}\ \emph {et~al.}(2014)\citenamefont {Yan},
  \citenamefont {Carretero-González}, \citenamefont {Frantzeskakis},
  \citenamefont {Kevrekidis}, \citenamefont {Proukakis},\ and\ \citenamefont
  {Spirn}}]{yan_14}%
  \BibitemOpen
  \bibfield  {author} {\bibinfo {author} {\bibfnamefont {D.}~\bibnamefont
  {Yan}}, \bibinfo {author} {\bibfnamefont {R.}~\bibnamefont
  {Carretero-González}}, \bibinfo {author} {\bibfnamefont {D.~J.}\
  \bibnamefont {Frantzeskakis}}, \bibinfo {author} {\bibfnamefont {P.~G.}\
  \bibnamefont {Kevrekidis}}, \bibinfo {author} {\bibfnamefont {N.~P.}\
  \bibnamefont {Proukakis}}, \ and\ \bibinfo {author} {\bibfnamefont
  {D.}~\bibnamefont {Spirn}},\ }\href {\doibase 10.1103/PhysRevA.89.043613}
  {\bibfield  {journal} {\bibinfo  {journal} {Phys. Rev. A}\ }\textbf {\bibinfo
  {volume} {89}},\ \bibinfo {pages} {043613} (\bibinfo {year}
  {2014})}\BibitemShut {NoStop}%
\bibitem [{\citenamefont {Mewes}\ \emph {et~al.}(1996)\citenamefont {Mewes},
  \citenamefont {Andrews}, \citenamefont {van Druten}, \citenamefont {Kurn},
  \citenamefont {Durfee}, \citenamefont {Townsend},\ and\ \citenamefont
  {Ketterle}}]{mewes_96}%
  \BibitemOpen
  \bibfield  {author} {\bibinfo {author} {\bibfnamefont {M.-O.}\ \bibnamefont
  {Mewes}}, \bibinfo {author} {\bibfnamefont {M.~R.}\ \bibnamefont {Andrews}},
  \bibinfo {author} {\bibfnamefont {N.~J.}\ \bibnamefont {van Druten}},
  \bibinfo {author} {\bibfnamefont {D.~M.}\ \bibnamefont {Kurn}}, \bibinfo
  {author} {\bibfnamefont {D.~S.}\ \bibnamefont {Durfee}}, \bibinfo {author}
  {\bibfnamefont {C.~G.}\ \bibnamefont {Townsend}}, \ and\ \bibinfo {author}
  {\bibfnamefont {W.}~\bibnamefont {Ketterle}},\ }\href {\doibase
  10.1103/PhysRevLett.77.988} {\bibfield  {journal} {\bibinfo  {journal} {Phys.
  Rev. Lett.}\ }\textbf {\bibinfo {volume} {77}},\ \bibinfo {pages} {988}
  (\bibinfo {year} {1996})}\BibitemShut {NoStop}%
\bibitem [{\citenamefont {Jin}\ \emph {et~al.}(1997)\citenamefont {Jin},
  \citenamefont {Matthews}, \citenamefont {Ensher}, \citenamefont {Wieman},\
  and\ \citenamefont {Cornell}}]{jin_97}%
  \BibitemOpen
  \bibfield  {author} {\bibinfo {author} {\bibfnamefont {D.~S.}\ \bibnamefont
  {Jin}}, \bibinfo {author} {\bibfnamefont {M.~R.}\ \bibnamefont {Matthews}},
  \bibinfo {author} {\bibfnamefont {J.~R.}\ \bibnamefont {Ensher}}, \bibinfo
  {author} {\bibfnamefont {C.~E.}\ \bibnamefont {Wieman}}, \ and\ \bibinfo
  {author} {\bibfnamefont {E.~A.}\ \bibnamefont {Cornell}},\ }\href {\doibase
  10.1103/PhysRevLett.78.764} {\bibfield  {journal} {\bibinfo  {journal} {Phys.
  Rev. Lett.}\ }\textbf {\bibinfo {volume} {78}},\ \bibinfo {pages} {764}
  (\bibinfo {year} {1997})}\BibitemShut {NoStop}%
\bibitem [{\citenamefont {Lin}\ \emph {et~al.}(2009)\citenamefont {Lin},
  \citenamefont {Compton}, \citenamefont {Perry}, \citenamefont {Phillips},
  \citenamefont {Porto},\ and\ \citenamefont {Spielman}}]{lin_09}%
  \BibitemOpen
  \bibfield  {author} {\bibinfo {author} {\bibfnamefont {Y.-J.}\ \bibnamefont
  {Lin}}, \bibinfo {author} {\bibfnamefont {R.~L.}\ \bibnamefont {Compton}},
  \bibinfo {author} {\bibfnamefont {A.~R.}\ \bibnamefont {Perry}}, \bibinfo
  {author} {\bibfnamefont {W.~D.}\ \bibnamefont {Phillips}}, \bibinfo {author}
  {\bibfnamefont {J.~V.}\ \bibnamefont {Porto}}, \ and\ \bibinfo {author}
  {\bibfnamefont {I.~B.}\ \bibnamefont {Spielman}},\ }\href@noop {} {\bibfield
  {journal} {\bibinfo  {journal} {Phys. Rev. Lett.}\ }\textbf {\bibinfo
  {volume} {102}},\ \bibinfo {pages} {130401} (\bibinfo {year}
  {2009})}\BibitemShut {NoStop}%
\bibitem [{\citenamefont {Hodby}\ \emph {et~al.}(2001)\citenamefont {Hodby},
  \citenamefont {Hechenblaikner}, \citenamefont {Hopkins}, \citenamefont
  {Marag\`o},\ and\ \citenamefont {Foot}}]{hodby_01}%
  \BibitemOpen
  \bibfield  {author} {\bibinfo {author} {\bibfnamefont {E.}~\bibnamefont
  {Hodby}}, \bibinfo {author} {\bibfnamefont {G.}~\bibnamefont
  {Hechenblaikner}}, \bibinfo {author} {\bibfnamefont {S.~A.}\ \bibnamefont
  {Hopkins}}, \bibinfo {author} {\bibfnamefont {O.~M.}\ \bibnamefont
  {Marag\`o}}, \ and\ \bibinfo {author} {\bibfnamefont {C.~J.}\ \bibnamefont
  {Foot}},\ }\href {\doibase 10.1103/PhysRevLett.88.010405} {\bibfield
  {journal} {\bibinfo  {journal} {Phys. Rev. Lett.}\ }\textbf {\bibinfo
  {volume} {88}},\ \bibinfo {pages} {010405} (\bibinfo {year}
  {2001})}\BibitemShut {NoStop}%
\bibitem [{\citenamefont {Khawaja}\ \emph {et~al.}(1999)\citenamefont
  {Khawaja}, \citenamefont {Pethick},\ and\ \citenamefont
  {Smith}}]{khawaja_99}%
  \BibitemOpen
  \bibfield  {author} {\bibinfo {author} {\bibfnamefont {U.~A.}\ \bibnamefont
  {Khawaja}}, \bibinfo {author} {\bibfnamefont {C.~J.}\ \bibnamefont
  {Pethick}}, \ and\ \bibinfo {author} {\bibfnamefont {H.}~\bibnamefont
  {Smith}},\ }\href {\doibase 10.1103/PhysRevA.60.1507} {\bibfield  {journal}
  {\bibinfo  {journal} {Phys. Rev. A}\ }\textbf {\bibinfo {volume} {60}},\
  \bibinfo {pages} {1507} (\bibinfo {year} {1999})}\BibitemShut {NoStop}%
\bibitem [{\citenamefont {Simula}\ \emph {et~al.}(2002)\citenamefont {Simula},
  \citenamefont {Virtanen},\ and\ \citenamefont {Salomaa}}]{simula_02}%
  \BibitemOpen
  \bibfield  {author} {\bibinfo {author} {\bibfnamefont {T.~P.}\ \bibnamefont
  {Simula}}, \bibinfo {author} {\bibfnamefont {S.~M.~M.}\ \bibnamefont
  {Virtanen}}, \ and\ \bibinfo {author} {\bibfnamefont {M.~M.}\ \bibnamefont
  {Salomaa}},\ }\href {\doibase 10.1103/PhysRevA.66.035601} {\bibfield
  {journal} {\bibinfo  {journal} {Phys. Rev. A}\ }\textbf {\bibinfo {volume}
  {66}},\ \bibinfo {pages} {035601} (\bibinfo {year} {2002})}\BibitemShut
  {NoStop}%
\bibitem [{\citenamefont {Pethick}\ and\ \citenamefont
  {Smith}(2008)}]{pethick_08}%
  \BibitemOpen
  \bibfield  {author} {\bibinfo {author} {\bibfnamefont {C.}~\bibnamefont
  {Pethick}}\ and\ \bibinfo {author} {\bibfnamefont {H.}~\bibnamefont
  {Smith}},\ }\href@noop {} {\emph {\bibinfo {title} {Bose-Einstein
  Condensation in Dilute Gases}}}\ (\bibinfo  {publisher} {Cambridge University
  Press},\ \bibinfo {year} {2008})\BibitemShut {NoStop}%
\bibitem [{\citenamefont {Muruganandam}\ and\ \citenamefont
  {Adhikari}(2009)}]{muruganandam_09}%
  \BibitemOpen
  \bibfield  {author} {\bibinfo {author} {\bibfnamefont {P.}~\bibnamefont
  {Muruganandam}}\ and\ \bibinfo {author} {\bibfnamefont {S.}~\bibnamefont
  {Adhikari}},\ }\href {\doibase https://doi.org/10.1016/j.cpc.2009.04.015}
  {\bibfield  {journal} {\bibinfo  {journal} {Comput. Phys. Commun.}\ }\textbf
  {\bibinfo {volume} {180}},\ \bibinfo {pages} {1888 } (\bibinfo {year}
  {2009})}\BibitemShut {NoStop}%
\bibitem [{\citenamefont {Vudragović}\ \emph {et~al.}(2012)\citenamefont
  {Vudragović}, \citenamefont {Vidanović}, \citenamefont {Balaž},
  \citenamefont {Muruganandam},\ and\ \citenamefont {Adhikari}}]{dusan_12}%
  \BibitemOpen
  \bibfield  {author} {\bibinfo {author} {\bibfnamefont {D.}~\bibnamefont
  {Vudragović}}, \bibinfo {author} {\bibfnamefont {I.}~\bibnamefont
  {Vidanović}}, \bibinfo {author} {\bibfnamefont {A.}~\bibnamefont {Balaž}},
  \bibinfo {author} {\bibfnamefont {P.}~\bibnamefont {Muruganandam}}, \ and\
  \bibinfo {author} {\bibfnamefont {S.~K.}\ \bibnamefont {Adhikari}},\ }\href
  {\doibase http://doi.org/10.1016/j.cpc.2012.03.022} {\bibfield  {journal}
  {\bibinfo  {journal} {Comput. Phys. Commun.}\ }\textbf {\bibinfo {volume}
  {183}},\ \bibinfo {pages} {2021 } (\bibinfo {year} {2012})}\BibitemShut
  {NoStop}%
\bibitem [{\citenamefont {Satarić}\ \emph {et~al.}(2016)\citenamefont
  {Satarić}, \citenamefont {Slavnić}, \citenamefont {Belić}, \citenamefont
  {Balaž}, \citenamefont {Muruganandam},\ and\ \citenamefont
  {Adhikari}}]{sataric_16}%
  \BibitemOpen
  \bibfield  {author} {\bibinfo {author} {\bibfnamefont {B.}~\bibnamefont
  {Satarić}}, \bibinfo {author} {\bibfnamefont {V.}~\bibnamefont {Slavnić}},
  \bibinfo {author} {\bibfnamefont {A.}~\bibnamefont {Belić}}, \bibinfo
  {author} {\bibfnamefont {A.}~\bibnamefont {Balaž}}, \bibinfo {author}
  {\bibfnamefont {P.}~\bibnamefont {Muruganandam}}, \ and\ \bibinfo {author}
  {\bibfnamefont {S.~K.}\ \bibnamefont {Adhikari}},\ }\href {\doibase
  https://doi.org/10.1016/j.cpc.2015.12.006} {\bibfield  {journal} {\bibinfo
  {journal} {Comput. Phys. Commun.}\ }\textbf {\bibinfo {volume} {200}},\
  \bibinfo {pages} {411 } (\bibinfo {year} {2016})}\BibitemShut {NoStop}%
\bibitem [{\citenamefont {Young-S.}\ \emph {et~al.}(2016)\citenamefont
  {Young-S.}, \citenamefont {Vudragović}, \citenamefont {Muruganandam},
  \citenamefont {Adhikari},\ and\ \citenamefont {Balaž}}]{young_16}%
  \BibitemOpen
  \bibfield  {author} {\bibinfo {author} {\bibfnamefont {L.~E.}\ \bibnamefont
  {Young-S.}}, \bibinfo {author} {\bibfnamefont {D.}~\bibnamefont
  {Vudragović}}, \bibinfo {author} {\bibfnamefont {P.}~\bibnamefont
  {Muruganandam}}, \bibinfo {author} {\bibfnamefont {S.~K.}\ \bibnamefont
  {Adhikari}}, \ and\ \bibinfo {author} {\bibfnamefont {A.}~\bibnamefont
  {Balaž}},\ }\href {\doibase https://doi.org/10.1016/j.cpc.2016.03.015}
  {\bibfield  {journal} {\bibinfo  {journal} {Comput. Phys. Commun.}\ }\textbf
  {\bibinfo {volume} {204}},\ \bibinfo {pages} {209 } (\bibinfo {year}
  {2016})}\BibitemShut {NoStop}%
\end{thebibliography}%
\bibliographystyle{apsrev4-1}

\end{document}